\documentclass{article}
\pdfoutput=1

\usepackage{arxiv}

\usepackage[utf8]{inputenc} 
\usepackage[T1]{fontenc}    
\usepackage{hyperref}       
\usepackage{url}            
\usepackage{booktabs}       
\usepackage{amsfonts}       
\usepackage{nicefrac}       
\usepackage{microtype}      
\usepackage{lipsum}		    

\IfFileExists{NJDnatbib.sty}{\usepackage[numbers,super]{NJDnatbib}}{\usepackage[numbers,super]{natbib}}

\usepackage{booktabs}
\usepackage{nicefrac}
\usepackage[utf8]{inputenc}
\usepackage[T1]{fontenc}
\usepackage{tikz}
\usepackage{amsmath}
\usepackage{amsthm}
\usepackage{amssymb}
\usepackage{graphicx}
\usepackage{multirow}

\usetikzlibrary{shapes.geometric,arrows,fit,matrix,positioning}
\tikzstyle{stepx} = [rectangle, rounded corners, minimum width=3cm, minimum height=1cm,text centered, draw=black, fill=red!30]
\tikzstyle{arrow} = [thick,->,>=stealth,text width=8.5cm]

\usepackage{todonotes}
\usepackage{rotating}

\title{Bayesian Model-Averaged Meta-Analysis in Medicine}

\author{
    František Bartoš               \\
	Department of Psychology       \\
	University of Amsterdam        \\
	Noord-Holland, The Netherlands \\
	\And
	Quentin F. Gronau              \\
	Department of Psychology       \\
	University of Amsterdam        \\
	Noord-Holland, The Netherlands \\
	\And
	Bram Timmers                   \\
	Department of Psychology       \\
	University of Amsterdam        \\
	Noord-Holland, The Netherlands \\
	\And
	Willem M. Otte \\
	Department of Pediatric Neurology, UMC Utrecht Brain Center\\
	University Medical Center Utrecht and Utrecht University \\
	Utrecht, The Netherlands \\
	\& \\
	Biomedical MR Imaging and Spectroscopy Group, Center for Image Sciences \\
	University Medical Center Utrecht and Utrecht University \\
	Utrecht, The Netherlands \\
	\And
	Alexander Ly                   \\
	Department of Psychology       \\
	University of Amsterdam        \\
	Noord-Holland, The Netherlands \\
	\& \\
	Centrum Wiskunde \& Informatica, Amsterdam \\
	Noord-Holland, The Netherlands \\
	\And
	Eric-Jan Wagenmakers           \\
	Department of Psychology       \\
	University of Amsterdam        \\
	Noord-Holland, The Netherlands \\
	}

\renewcommand{\shorttitle}{Bayesian Model-Averaged Meta-Analysis in Medicine}

\begin{document}
\maketitle

\begin{abstract}
We outline a Bayesian model-averaged meta-analysis for standardized mean differences in order to quantify evidence for both treatment effectiveness $\delta$ and across-study heterogeneity $\tau$. We construct four competing models by orthogonally combining two present-absent assumptions, one for the treatment effect and one for across-study heterogeneity. To inform the choice of prior distributions for the model parameters, we used 50\% of the Cochrane Database of Systematic Reviews to specify rival prior distributions for $\delta$ and $\tau$. The relative predictive performance of the competing models and rival prior distributions was assessed using the remaining 50\% of the Cochrane Database. On average, $\mathcal{H}_1^r$ -- the model that assumes the presence of a treatment effect as well as across-study heterogeneity -- outpredicted the other models, but not by a large margin. Within $\mathcal{H}_1^r$, predictive adequacy was relatively constant across the rival prior distributions. We propose specific empirical prior distributions, both for the field in general and for each of 46 specific medical subdisciplines. An example from oral health demonstrates how the proposed prior distributions can be used to conduct a Bayesian model-averaged meta-analysis in the open-source software \texttt{R} and JASP. The preregistered analysis plan is available at \url{https://osf.io/zs3df/}.
\end{abstract}


\section{Introduction}

Following Karl Pearson's first quantitative synthesis of clinical trials in 1904, meta-analysis gradually established itself as an irreplaceable method for statistics in medicine.\cite{o2007historical} However, over a century later meta-analysis still presents formidable statistical challenges to medical practitioners, especially when the number of primary studies is low. In this case the estimation of across-study heterogeneity (i.e., across-study standard deviation) $\tau$ is problematic \cite{brockwell2001comparison, inthout2014hartung, gonnermann2015no}; moreover, these problematic $\tau$ estimates may subsequently distort the estimates of the overall treatment effect size $\delta$.\cite{brockwell2001comparison,berkey1995random} The practical relevance of the small sample challenge is underscored by the fact that the median number of studies in a meta-analysis from the Cochrane Database of Systematic Reviews (CDSR) is only 3, with an interquartile range from 2 to 6.\cite{Davey2011characteristics}

One statistical method that has been proposed to address the small sample challenge is \emph{Bayesian estimation}, either with weakly informative prior distributions, \cite{williams2018bayesian, higgins2009re, chung2013avoiding} predictive prior distributions based on pseudo-data,\cite{rhodes2016implementing} or prior distributions informed by earlier studies.\cite{higgins1996borrowing} These Bayesian techniques are well suited to estimate the model parameters when the data are scarce; however, by assigning continuous prior distributions to $\delta$ and $\tau$, these estimation techniques implicitly assume that the treatment is effective and the studies are not homogeneous.\footnote{Under a continuous prior distribution, the prior probability of any particular value (such as $\delta=0$, which represent the proposition that the treatment is ineffective) is zero.} In order to validate these strong assumptions we may adopt the framework of \emph{Bayesian testing}. Developed in the second half of the 1930s by Sir Harold Jeffreys,\cite{Jeffreys1935,Jeffreys1939} the Bayesian testing framework seeks to grade the evidence that the data provide for or against a specific value of interest such as $\delta=0$ and $\tau=0$ which corresponds to the null model of no effect and the fixed-effect model, respectively. Jeffreys argued that the testing question logically precedes the estimation question, and that more complex models (e.g., the models used for estimation, where $\delta$ and $\tau$ are free parameters) ought to be adopted only after the data provide positive evidence in their favor: ``Until such evidence is actually produced the simpler hypothesis holds the field; the onus of proof is always on the advocate of the more complicated hypothesis.''\cite[p. 252]{Jeffreys1937SI}

In the context of meta-analysis, Jeffreys's statistical philosophy demands that we acknowledge not only the uncertainty in the parameter values given a specific model, but also the uncertainty in the underlying models to which the parameters belong. Both types of uncertainty can be assessed and updated using a procedure known as Bayesian model-averaged (BMA) meta-analysis.\footnote{A different approach is Bayesian model selection based on posterior
(out-of-sample) predictive performance such as DIC/WAIC/LOO.\cite{yao2018using, VehtariEtAl2017} However, these approaches are unable to provide compelling support in favor of simple models.\cite{GronauWagenmakers2019LOO1, GronauWagenmakers2019LOO2}} The BMA procedure applies different meta-analytic models to the data simultaneously, and draws inferences by taking into account all models, with their impact determined by their predictive performance for the observed data.\cite{GronauEtAl2017PowerPose, HinneEtAl2020, gronau2020primer}

As in other applications of Bayesian statistics, BMA requires that all parameters are assigned prior distributions. However, in contrast to Bayesian estimation, Bayesian testing does not permit the specification of vague or ``uninformative'' prior distributions on the parameters of interest. Vague prior distributions assign most prior mass to implausibly large values, resulting in poor predictive performance.\cite{Jeffreys1939, gronau2020primer,KassRaftery1995} In BMA, the relative impact of the models is determined by their predictive performance, and predictive performance in turn is determined partly by the prior distribution on the model parameters. In objective Bayesian statistics\cite{ConsonniEtAl2018} so-called \emph{default prior distributions} have been proposed; these default distributions meet a list of desiderata \cite{BayarriEtAl2012} and are intended for general use in testing. In contrast to this work, here we seek to construct and compare different prior distributions based on existing medical knowledge. \cite{turner2015predictive,pullenayegum2011informed,rhodes2015predictive} Specifically, we propose \emph{empirical prior distributions} for $\delta$ and $\tau$ as applied to meta-analyses of continuous outcomes in medicine. To this aim we first used 50\% of CDSR to develop candidate prior distributions and then used the remaining 50\% of CDSR to evaluate their predictive accuracy and that of the associated models.

Below we first outline the BMA approach to meta-analyses and then present the results of a preregistered analysis procedure to obtain and assess empirical prior distributions for $\delta$ and $\tau$ for the medical field as a whole. Next we propose empirical prior distributions for the 46 specific medical subdisciplines defined by CDSR. Finally we demonstrate with a concrete example how our results can be applied in practice using the open-source statistical programs \texttt{R}\cite{R} and JASP.\cite{JASP15}

\section{Bayesian Model-Averaged Meta-Analysis}
The standard Bayesian random-effects meta-analysis assumes that a latent individual study effect $\theta_i$ is drawn from a Gaussian group-level distribution with mean treatment effect $\delta$ and between-study heterogeneity $\tau$.\cite{SpiegelhalterEtAl1994,SuttonAbrams2001} Inference then concerns the posterior distributions for $\delta$ and $\tau$. This estimation approach allows researchers to answer important questions such as ``given that the treatment effect is nonzero, how large is it?''\footnote{More specific versions of this generic question are ``given that the treatment effect is nonzero, is it positive or negative?'' and ``given that the treatment effect is nonzero, what is the posterior probability that it falls in the interval from $a$ to $b$?''} and ``given that there is between-study heterogeneity, how large is it?'' Because the standard model assumes that the effect is nonzero, it cannot address the arguably more fundamental questions that involve a hypothesis test,\cite{HaafEtAl2019Nature,Jeffreys1961} such as ``how strong is the evidence in favor of the presence or absence of a treatment effect?'' and ``how strong is the evidence in favor of between-study heterogeneity (between-study standard deviation) versus homogeneity?''\cite[p. 274]{Fisher1928} Here we outline a BMA approach that allows for both hypothesis testing and parameter estimation in a single statistical framework.\cite{RoverEtAl2019}

Our generic meta-analysis setup\cite{GronauEtAl2017PowerPose,HinneEtAl2020,LandyEtAl2020,ScheibehenneEtAl2017Reply} (for the conceptual basis see Jeffreys)\cite[p. 276-277 and p. 296]{Jeffreys1939} consists of the following four qualitatively different candidate hypotheses\footnote{The terms `hypothesis' and `model' are used interchangeably.}:
\begin{enumerate}
    \item the fixed-effect null hypothesis $\mathcal{H}_0^f$  : $ \delta = 0$ , $\tau = 0$;
    \item the fixed-effect alternative hypothesis $\mathcal{H}_{1}^f$  : $ \delta \sim g(	\cdot)$ , $\tau = 0$;
    \item the random-effects null hypothesis $\mathcal{H}_0^r$ : $\delta = 0$, $\tau \sim h(\cdot)$;
    \item the random-effects alternative hypothesis $\mathcal{H}_{1}^r$  : $\delta \sim g(	\cdot)$ , $\tau \sim h(\cdot)$,
\end{enumerate}
where $\delta$ represents the group-level mean treatment effect, $\tau$ represents the between-study standard deviation (i.e., the treatment heterogeneity), and $g(\cdot)$ and $h(\cdot)$ represent prior distributions that quantify the uncertainty about $\delta$ and $\tau$, respectively. The four prior probabilities of the rival hypotheses are denoted by $p(\mathcal{H}_0^f)$, $p(\mathcal{H}_{1}^f)$, $p(\mathcal{H}_0^r)$, and $p(\mathcal{H}_{1}^r)$; these may or may not be set to $\nicefrac{1}{4}$, reflecting a position of prior equipoise. The main advantage of this framework is that it does not fully commit to any single model on purely \emph{a priori} grounds. Although in many situations the random-effects alternative hypothesis $\mathcal{H}_{1}^r$ is an attractive option, it may be less appropriate when the number of studies is small; in addition, as mentioned above, $\mathcal{H}_{1}^r$ assumes the effect to be present, whereas assessing the degree to which the data undercut or support this assumption may often be one of the primary inferential goals. 

In our framework, after specifying the requisite prior distributions $g(\cdot)$ and $h(\cdot)$, the data drive an update from prior to posterior model probabilities, and pertinent conclusions are then drawn using BMA.\cite{Jevons18741913,HoetingEtAl1999} Specifically, the posterior odds of an effect being present, based on observed data $y$, is the ratio of the sum of posterior model probabilities for $\mathcal{H}_{1}^f$ and $\mathcal{H}_{1}^r$ over the sum of posterior model probabilities for $\mathcal{H}_{0}^f$ and $\mathcal{H}_{0}^r$:
\begin{equation*}
\text{Posterior odds for treatment effect} = \frac{p(\mathcal{H}_{1}^f \mid y) + p(\mathcal{H}_{1}^r \mid y)}{p(\mathcal{H}_{0}^f \mid y) + p(\mathcal{H}_{0}^r \mid y)}.    
\end{equation*}
In model-averaging terms, this quantity is referred to as the \emph{posterior inclusion odds}, as it refers to the post-data odds of `including' the effect size parameter $\delta$. As a measure of evidence, one may consider the change, brought about by the data, from prior inclusion odds to posterior inclusion odds. This change is known as the \emph{Bayes factor}\cite{KassRaftery1995,Jeffreys1961,EtzWagenmakers2017}:
\begin{equation}
\label{eq:BF_treatment}
\underbrace{\text{BF}_{10}}_{\substack{\text{ Inclusion Bayes factor}\\{\text{for treatment effect}}}} = \;\;\;\;
\underbrace{ \frac{p(\mathcal{H}_{1}^f \mid y) + p(\mathcal{H}_{1}^r \mid y)}{p(\mathcal{H}_{0}^f \mid y) + p(\mathcal{H}_{0}^r \mid y)}}_{\substack{\text{Posterior inclusion odds}\\{\text{for treatment effect}}}} \;\;\;\; \mbox{\Huge /} \;\;\, \underbrace{ \frac{p(\mathcal{H}_{1}^f) + p(\mathcal{H}_{1}^r)}{p(\mathcal{H}_{0}^f) + p(\mathcal{H}_{0}^r)}}_{\substack{\text{Prior inclusion odds}\\{\text{for treatment effect}}}}.
\end{equation}
One may similarly assess the posterior odds for the presence of heterogeneity by contrasting $\mathcal{H}_{0}^r$ and $\mathcal{H}_{1}^r$ versus $\mathcal{H}_{0}^f$ and $\mathcal{H}_{1}^f$:
\begin{equation*}
\text{Posterior odds for treatment heterogeneity} = \frac{p(\mathcal{H}_{0}^r \mid y) + p(\mathcal{H}_{1}^r \mid y)}{p(\mathcal{H}_{0}^f \mid y) + p(\mathcal{H}_{1}^f \mid y)},    
\end{equation*}
or one may quantify evidence by the change from prior to posterior inclusion odds:
\begin{equation}
\label{eq:BF_heterogeneity}
\underbrace{\text{BF}_{\text{rf}}}_{\substack{\text{ Inclusion Bayes factor}\\{\text{for treatment heterogeneity}}}} = \;\;\;\;
\underbrace{ \frac{p(\mathcal{H}_{0}^r \mid y) + p(\mathcal{H}_{1}^r \mid y)}{p(\mathcal{H}_{0}^f \mid y) + p(\mathcal{H}_{1}^f \mid y)}}_{\substack{\text{Posterior inclusion odds}\\{\text{for treatment heterogeneity}}}} \;\;\;\; \mbox{\Huge /} \underbrace{\frac{p(\mathcal{H}_{0}^r) + p(\mathcal{H}_{1}^r)}{p(\mathcal{H}_{0}^f) + p(\mathcal{H}_{1}^f)}}_{\substack{\text{Prior inclusion odds}\\{\text{for treatment heterogeneity}}}}.
\end{equation}

An attractive feature of this framework is that it allows a graceful data-driven transition from an emphasis on fixed-effect models to random-effects models; with only few studies available, the fixed-effect models likely outpredict the random-effects models and therefore receive more weight. But as studies accumulate, and it becomes increasingly apparent that the treatment effect is indeed random, the influence of the random-effects models will wax and of the fixed-effect models will wane, until inference is dominated by the random-effects models. In addition, the Bayesian framework allows researchers to monitor the evidence as studies accumulate, without the need or want of corrections for optional stopping.\cite{BergerWolpert1988} This is particularly relevant as the accumulation of studies is usually not under the control of a central agency, and the stopping rule is ill-defined. \cite{ter_schure_accumulation_2019}

Although theoretically promising, the practical challenge for our BMA meta-analysis is to determine appropriate prior distributions for $\delta$ and $\tau$. Prior distributions that are too wide will waste prior mass on highly implausible parameter values, thus incurring a penalty for complexity that could have been circumvented by applying a more reasonably peaked prior distribution. On the other hand, prior distributions that are too narrow represent a highly risky bet; if the effect is not exactly where the peaked prior distribution guesses it to be, the model will incur a hefty penalty for predicting the data poorly, a penalty that could have been circumvented by reasonably widening the prior distribution. There is no principled way around this dilemma: Bayes' rule dictates that evidence is quantified by predictive success, and \emph{predictions} follow from the \emph{prior} predictive distributions.\cite{Evans2015} Thus, when the goal is to quantify evidence, the prior distributions warrant careful consideration.\citep{GronauEtAl2020Informed,LambertEtAl2005,OHaganEtAl2006,OHagan2019,TurnerEtAl2012}

Fortunately, the framework presented here contains only two key parameters, $\delta$ and $\tau$; moreover, a large clinical literature is available to help guide the specification of reasonable prior distributions. Our goal in this work is to use meta-analyses from the CDSR to create a series of informed prior distributions for both the effect size parameter $\delta$ and between-study variance parameter $\tau$.\cite{higgins2009re,turner2015predictive,pullenayegum2011informed,rhodes2015predictive} We will then assess the predictive adequacy of the various models in conjunction with the prior distributions on a hold-out validation set. 

\section{Candidate Prior Distributions}

We developed and assessed prior distributions for the $\delta$ and $\tau$ parameters suitable for BMA of continuous outcomes using data from CDSR.\footnote{We identified systematic reviews in the CDSR through PubMed, limiting the period to Jan 2000 -- May 2020. For that we used the NCBI's EUtils API with the following query: ``Cochrane Database Syst Rev''[journal] AND (``2000/01/01''[PDAT]: ``2020/05/31''[PDAT]). For each review, we downloaded the XML meta-analysis table file (rm5-format) associated with the review's latest version. We extracted the tables with continuous outcomes (i.e., MD and SMD) from these rm5-files with a custom PHP script. We labeled the tables with the Cochrane Review Group's taxonomy list for subfield analysis.} In the remainder of this work we adopt the terminology of Higgins et al.\cite{higgins2019cochrane}: individual meta-analyses included in each Cochrane review are referred to as `comparisons' and individual studies included in a comparison are referred to as `studies'. All of the results were conducted using Cohen's $d$ standardized mean differences (SMD). The analyses presented in this section were executed in accordance with a preregistration protocol (\url{https://osf.io/zs3df/}) unless explicitly mentioned otherwise.

In order to assess the predictive adequacy of the various prior distributions and models, we first randomly partitioned the data of the Cochrane reviews in a training and test set. The training set consisted of 3,092 comparisons with a total number of 23,333 individual studies, and the test set consisted of 3,091 comparisons with a total number of 22,117 individual studies. We used the training set to develop prior distributions for the $\delta$ and $\tau$ parameters and then assessed predictive accuracy using the test set.

\subsection{Developing Prior Distributions Based on the Training Set}
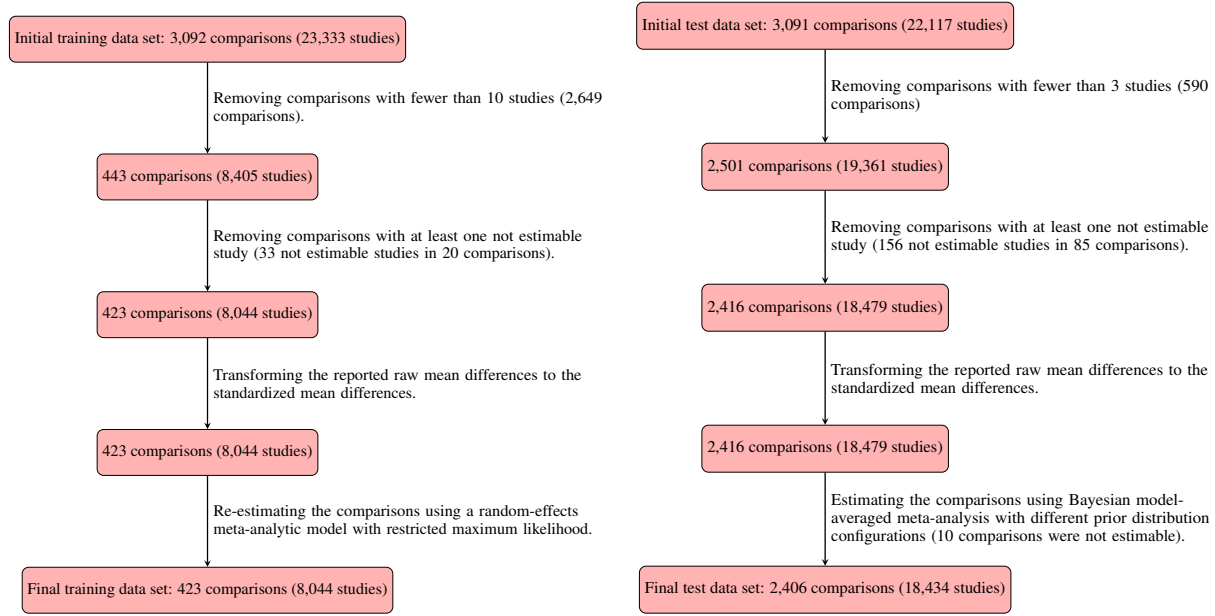
\begin{figure}[h!]
\centering
{
    \resizebox{.49 \textwidth}{!}{
        \begin{tikzpicture}[node distance=3cm]
            \node (step0) [stepx] {Initial training data set: 3,092 comparisons (23,333 studies)};
            \node (step1) [stepx, below of=step0] {443 comparisons (8,405 studies)};
            \node (step2) [stepx, below of=step1] {423 comparisons (8,044 studies)};
            \node (step3) [stepx, below of=step2] {423 comparisons (8,044 studies)};
            \node (step4) [stepx, below of=step3] {Final training data set: 423 comparisons (8,044 studies)};
            
            \draw [arrow] (step0) -- node[anchor=west] {Removing comparisons with fewer than 10 studies (2,649 comparisons).} (step1);
            \draw [arrow] (step1) -- node[anchor=west] {Removing comparisons with at least one not estimable study (33 not estimable studies in 20 comparisons).}(step2);
            \draw [arrow] (step2) -- node[anchor=west] {Transforming the reported raw mean differences to the standardized mean differences.}(step3);
            \draw [arrow] (step3) -- node[anchor=west] {Re-estimating the comparisons using a random-effects meta-analytic model with restricted maximum likelihood.}(step4);
        \end{tikzpicture}
    }
}
\hfill 
{
    \resizebox{.49 \textwidth}{!}{
        \begin{tikzpicture}[node distance=3cm]
            \node (step0) [stepx] {Initial test data set: 3,091 comparisons (22,117 studies)};
            \node (step1) [stepx, below of=step0] {2,501 comparisons (19,361 studies)};
            \node (step2) [stepx, below of=step1] {2,416 comparisons (18,479 studies)};
            \node (step3) [stepx, below of=step2] {2,416 comparisons (18,479 studies)};
            \node (step4) [stepx, below of=step3] {Final test data set: 2,406 comparisons (18,434 studies)};
            
            \draw [arrow] (step0) -- node[anchor=west] {Removing comparisons with fewer than 3 studies (590 comparisons)} (step1);
            \draw [arrow] (step1) -- node[anchor=west] {Removing comparisons with at least one not estimable study (156 not estimable studies in 85 comparisons).}(step2);
            \draw [arrow] (step2) -- node[anchor=west] {Transforming the reported raw mean differences to the standardized mean differences.}(step3);
            \draw [arrow] (step3) -- node[anchor=west] {Estimating the comparisons using Bayesian model-averaged meta-analysis with different prior distribution configurations (10 comparisons were not estimable).}(step4);
        \end{tikzpicture}
    }
}
\caption{Flowchart of the study selection procedure and data processing steps for the training data set (left) and the test data set (right).}
\label{fig:flowchart}
\end{figure}

\begin{figure}[h]
\begin{tabular}{cc}
    \begin{minipage}{.50 \textwidth}
    \includegraphics[width=1\textwidth]{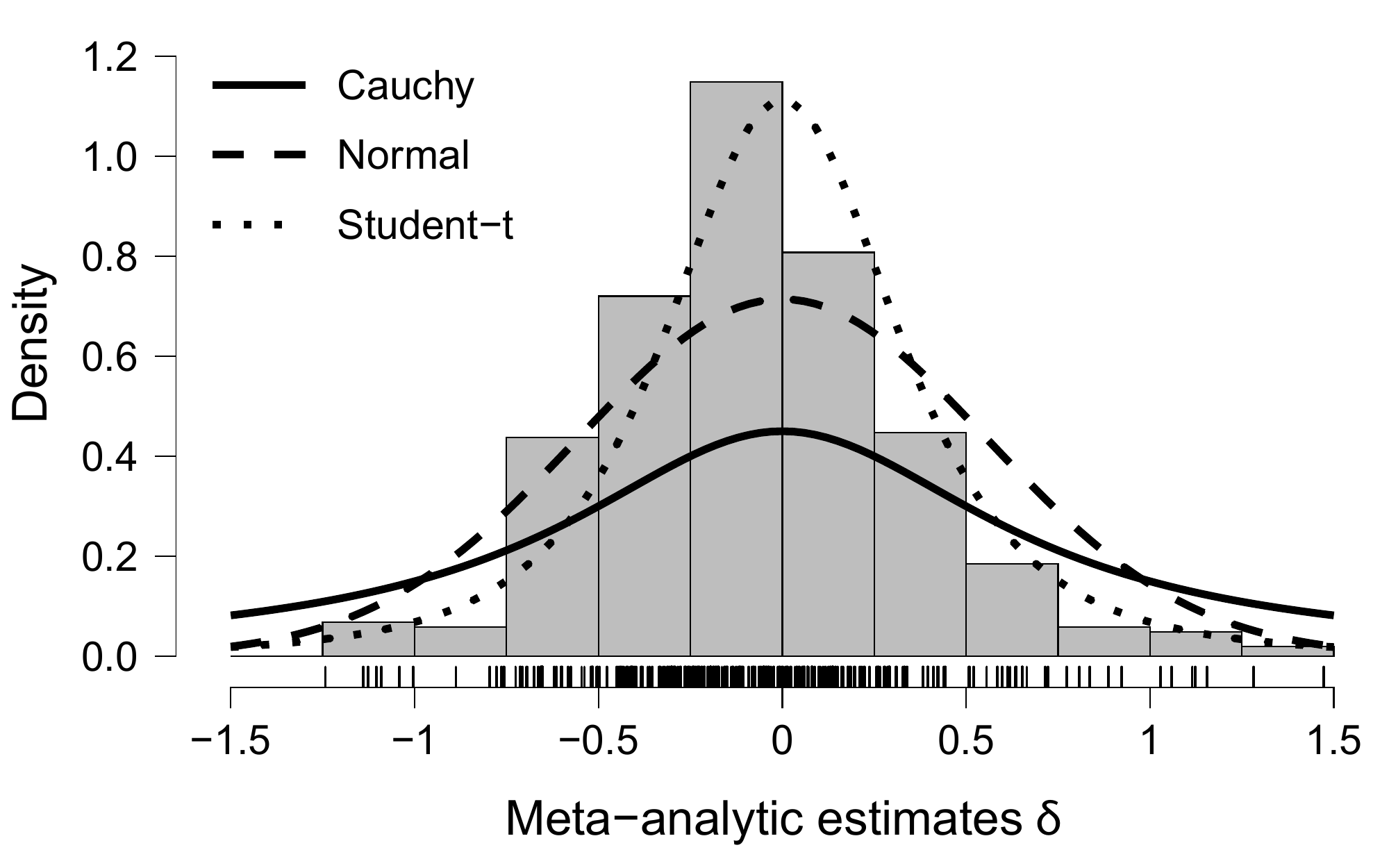}
    \end{minipage}  & %
    \begin{minipage}{.50 \textwidth}
    \includegraphics[width=1\textwidth]{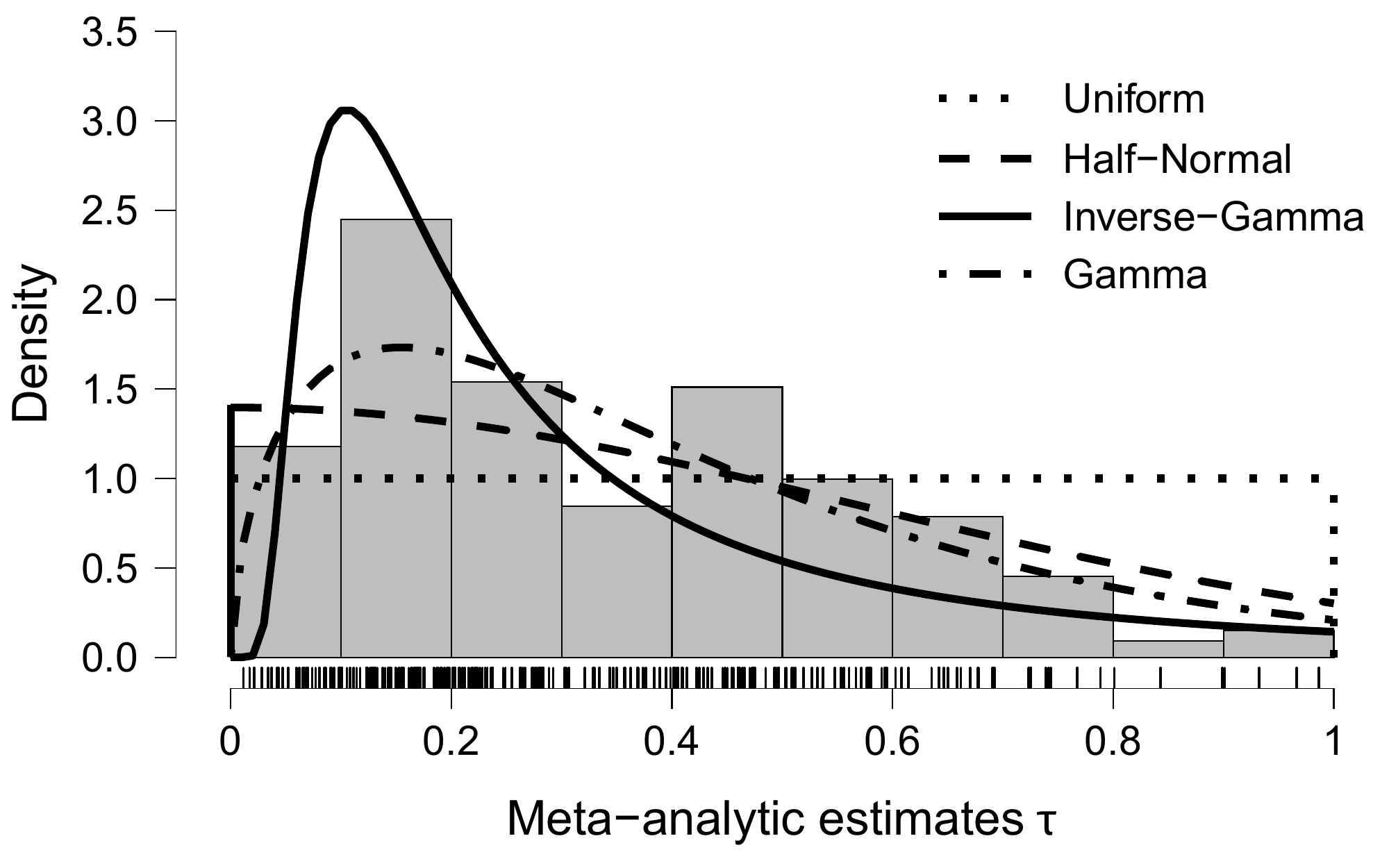}
    \end{minipage}  %
\end{tabular}
\caption{Frequentist effect sizes estimates and candidate prior distributions from the training data set. Histogram and tick marks display the estimated effect size estimates (left) and between-study standard deviation estimates (right), whereas lines represent three associated candidate prior distributions for the population effect size parameter $\delta$ (left) and four candidate prior distributions for the population between-study standard deviation $\tau$ (right; see Table~\ref{tab:prior-training-dataset}). Twelve effect sizes outside of the $\pm1.5$ range are not shown and twenty-four $\tau$ estimates larger than 1 and sixty-eight $\tau$ estimates lower than 0.01 are not shown.}
\label{fig:prior-training-dataset}
\end{figure}

Left panel of Figure~\ref{fig:flowchart} outlines the data processing steps performed on the training set (further details are provided in the preregistration protocol, \url{https://osf.io/zs3df/}). First, in order to ensure that the training set yields estimates of $\tau$ that form a reliable basis for the construction of a prior distribution, we excluded comparisons with fewer than 10 studies. Second, we excluded comparisons for which at least one individual study was reported by the authors of the review to be ``non-estimable'' (i.e., the effect size of the original study could not be retrieved). Third, we transformed the reported raw mean differences to the SMD using the \texttt{metafor} \texttt{R} package.\cite{viechtbauer2010metafor} Fourth, to ensure high consistency of the meta-analytic estimates, we re-estimated all comparisons using a frequentist random-effects meta-analytic model with restricted maximum likelihood estimator using the \texttt{metafor} \texttt{R} package.\cite{viechtbauer2010metafor} These steps resulted in a final training set featuring 423 comparisons containing a total of 8,044 individual studies. The histograms and tick marks in Figure~\ref{fig:prior-training-dataset} display the $\delta$ and $\tau$ estimates from each comparison in the training set.\footnote{As specified in the preregistration protocol, we assumed that $\tau$ estimates lower than $0.01$ are representative of $\mathcal{H}_\cdot^f: \tau = 0$, and therefore these estimates were not used to determine candidate prior distributions for $\tau$.}

To develop candidate prior distributions for parameters $\delta$ and $\tau$, we used the maximum likelihood estimator implemented in the \texttt{fitdistrplus} \texttt{R} package\cite{fitdistrplus} to fit several distributions to the frequentist meta-analytic estimates from the training set. For the $\delta$ parameter, we considered normal and Student's $t$ distributions fitted to the training set and compared them to an uninformed Cauchy distribution with scale $\nicefrac{1}{\sqrt{2}}$ (a default choice in the field of psychology).\cite{morey2015bayesfactor} For the $\tau$ parameter, we considered half-normal, inverse-gamma, and gamma distributions fitted to the training set and compared them to an uninformed uniform distribution on the range from 0 to 1.\cite{LambertEtAl2005} The resulting distributions are summarized in Table~\ref{tab:prior-training-dataset} and their fit to the training set is visualized in Figure~\ref{fig:prior-training-dataset}.

\begin{table*}[h]
  \centering
  \caption{Candidate prior distributions for the $\delta$ and $\tau$ parameters as obtained from the training set. The inverse-gamma and gamma distributions follow the shape and scale parametrization and the Studen-t distributions follow the location, scale, and degrees of freedom parametrization. See Figure~\ref{fig:prior-training-dataset}.}
  \label{tab:prior-training-dataset}
    \begin{tabular}{ll}
    \toprule
    Parameter $\delta$                                     & Parameter $\tau$ \\ \midrule
    $\delta \sim \text{Cauchy}(0,\nicefrac{1}{\sqrt{2}})$  & $\tau \sim \mathcal{U}(0, 1)$               \\
    $\delta \sim \mathcal{N}(0, 0.56^2)$                   & $\tau \sim \mathcal{N}_+(0, 0.57^2)$        \\
    $\delta \sim \mathcal{T}(0, 0.33, 3)$                  & $\tau \sim \text{Inv-Gamma}(1.26, 0.24)$  \\
                                                           & $\tau \sim \text{Gamma}(1.59, 0.26)$        \\ 
    \bottomrule
  \end{tabular}
\end{table*}

\subsection{Assessing Prior Distributions Based on the Test Set}

Right panel of Figure~\ref{fig:flowchart} outlines the data processing steps performed on the test set. Similarly to the training set, we removed non-estimable comparisons  and transformed all effect sizes to SMD. However, in contrast to the training set, we retained all comparisons that feature at least 3 studies: there is no reason to limit the assessment of predictive performance to comparisons with at least 10 studies. These data processing steps resulted in a final test set consisting of 2,416 comparisons containing a total of 18,479 individual studies. The median number of studies in a comparison was 5 with an interquartile range from 3 to 9.

For each possible pair of candidate prior distributions depicted in Table~\ref{tab:prior-training-dataset},\footnote{We failed to estimate all BMA meta-analytic models for ten comparisons due to large outliers.} we computed posterior model probabilities and model-averaged Bayes factors with the \texttt{metaBMA} \texttt{R} package,\cite{heck2017metabma} which uses numerical integration and bridge sampling.\cite{GronauEtAl2017BSTutorial, GronauEtAl2020JSS, MengWong1996}

\subsubsection{Performance of Prior Distribution Configurations Under $\mathcal{H}^r_1$}

In the first analysis we evaluate the predictive performance associated with the different prior distribution configurations as implemented in $\mathcal{H}_1^r$, that is the random-effects model that allows both $\delta$ and $\tau$ to be estimated from the data. Specifically, under $\mathcal{H}_1^r$ there are $3 \times 4 = 12$ prior configurations and each is viewed as a model yielding predictions. The prior probability of each prior configuration is $\nicefrac{1}{12} \approx 0.083$ and the predictive accuracy of each prior configuration is assessed with the 2406 comparisons from the test set. Table~\ref{tab:ranking_prior_configurations} lists the 12 different prior configurations and summarizes the number of times their posterior probability was ranked $1,2,\ldots,12$. The results show that informed configurations generally outperformed the uninformed configurations (i.e., the $\text{Cauchy}(0,\nicefrac{1}{\sqrt{2}})$ distribution on $\delta$ and the $\mathcal{U}(0, 1)$ distribution on $\tau$). The worst ranking performance was obtained with prior configuration 1 (i.e., uniformed distributions for both $\delta$ and $\tau$). Prior configurations 2, 3, 4, 5, and 9 feature an uninformed prior distribution on either $\delta$ or $\tau$, and also did not perform well in terms of posterior rankings. The same holds for prior distribution configurations with the half-normal prior distribution for the $\tau$ parameter (i.e., prior configurations 2, 6, and 10). The best performing prior distribution configurations (i.e., numbers 7, 11, and 12) used more data-driven prior distributions for both $\delta$ (i.e., fitted normal or $t$-distributions) and $\tau$ (i.e., fitted inverse-gamma or gamma).

\begin{sidewaystable}[ph!]
  \caption{Ranking totals for each prior configuration based on the 2406 comparisons in the test set. The numbers indicate how many times a specific prior configuration attained a specific posterior probability rank amongst the 12 possible prior configurations. Rank `1' represents the best performance. Note that these rankings are conditional on assuming the meta-analytic model $\mathcal{H}_1^r$ (i.e., the posterior probabilities of the other meta-analytic models are not considered).}
  \label{tab:ranking_prior_configurations}
  \tiny
  \resizebox{\textwidth}{!}{
  \begin{tabular}{@{}lllllllllllllll@{}}
    \toprule
    \multicolumn{2}{l}{} & \multicolumn{12}{c}{Rank} & \multicolumn{1}{l}{}\\
Prior $\delta$                                            & Prior $\tau$                             & 1 & 2 & 3 & 4 & 5 & 6 & 7 & 8 & 9 & 10 & 11 & 12 \\ \midrule 
1.  $\delta \sim \text{Cauchy}(0,\nicefrac{1}{\sqrt{2}})$ & $\tau \sim \mathcal{U}(0, 1)$            & 67 & 74 & 35 & 92 & 51 & 56 & 128 & 89 & 91 & 111 & 128 & 1484  \\  
2.  $\delta \sim \text{Cauchy}(0,\nicefrac{1}{\sqrt{2}})$ & $\tau \sim \mathcal{N}_+(0, 0.57^2)$     & 7 & 54 & 39 & 38 & 62 & 82 & 103 & 170 & 134 & 327 & 1390 & 0   \\  
3.  $\delta \sim \text{Cauchy}(0,\nicefrac{1}{\sqrt{2}})$ & $\tau \sim \text{Inv-Gamma}(1.26, 0.24)$ & 54 & 17 & 46 & 64 & 35 & 55 & 109 & 455 & 706 & 163 & 133 & 569  \\   
4.  $\delta \sim \text{Cauchy}(0,\nicefrac{1}{\sqrt{2}})$ & $\tau \sim \text{Gamma}(1.59, 0.26)$     & 9 & 23 & 41 & 51 & 73 & 55 & 62 & 84 & 493 & 1249 & 261 & 5  \\  
5.  $\delta \sim \mathcal{N}(0, 0.56^2)$                  & $\tau \sim \mathcal{U}(0, 1)$            & 282 & 180 & 50 & 139 & 75 & 101 & 155 & 853 & 443 & 59 & 39 & 30  \\  
6.  $\delta \sim \mathcal{N}(0, 0.56^2)$                  & $\tau \sim \mathcal{N}_+(0, 0.57^2)$     & 8 & 110 & 350 & 217 & 253 & 470 & 933 & 38 & 13 & 9 & 5 & 0  \\  
7.  $\delta \sim \mathcal{N}(0, 0.56^2)$                  & $\tau \sim \text{Inv-Gamma}(1.26, 0.24)$ & 247 & 226 & 211 & 724 & 143 & 120 & 106 & 229 & 139 & 191 & 66 & 4  \\  
8.  $\delta \sim \mathcal{N}(0, 0.56^2)$                  & $\tau \sim \text{Gamma}(1.59, 0.26)$     & 123 & 189 & 128 & 230 & 677 & 808 & 196 & 22 & 15 & 6 & 8 & 4  \\  
9.  $\delta \sim \mathcal{T}(0, 0.33, 3)$                 & $\tau \sim \mathcal{U}(0, 1)$            & 227 & 162 & 79 & 283 & 549 & 303 & 304 & 166 & 72 & 55 & 94 & 112  \\  
10. $\delta \sim \mathcal{T}(0, 0.33, 3)$                 & $\tau \sim \mathcal{N}_+(0, 0.57^2)$     & 30 & 197 & 1051 & 257 & 233 & 199 & 111 & 93 & 99 & 83 & 53 & 0  \\  
11. $\delta \sim \mathcal{T}(0, 0.33, 3)$                 & $\tau \sim \text{Inv-Gamma}(1.26, 0.24)$ & 1122 & 207 & 111 & 137 & 60 & 47 & 101 & 140 & 92 & 52 & 155 & 182  \\  
12. $\delta \sim \mathcal{T}(0, 0.33, 3)$                 & $\tau \sim \text{Gamma}(1.59, 0.26)$     & 230 & 967 & 265 & 174 & 195 & 110 & 98 & 67 & 109 & 101 & 74 & 16  \\
  \\ \bottomrule
  \end{tabular}
  }
\end{sidewaystable} 

\begin{figure*}[h]
\centering
    \includegraphics[width=0.6\textwidth]{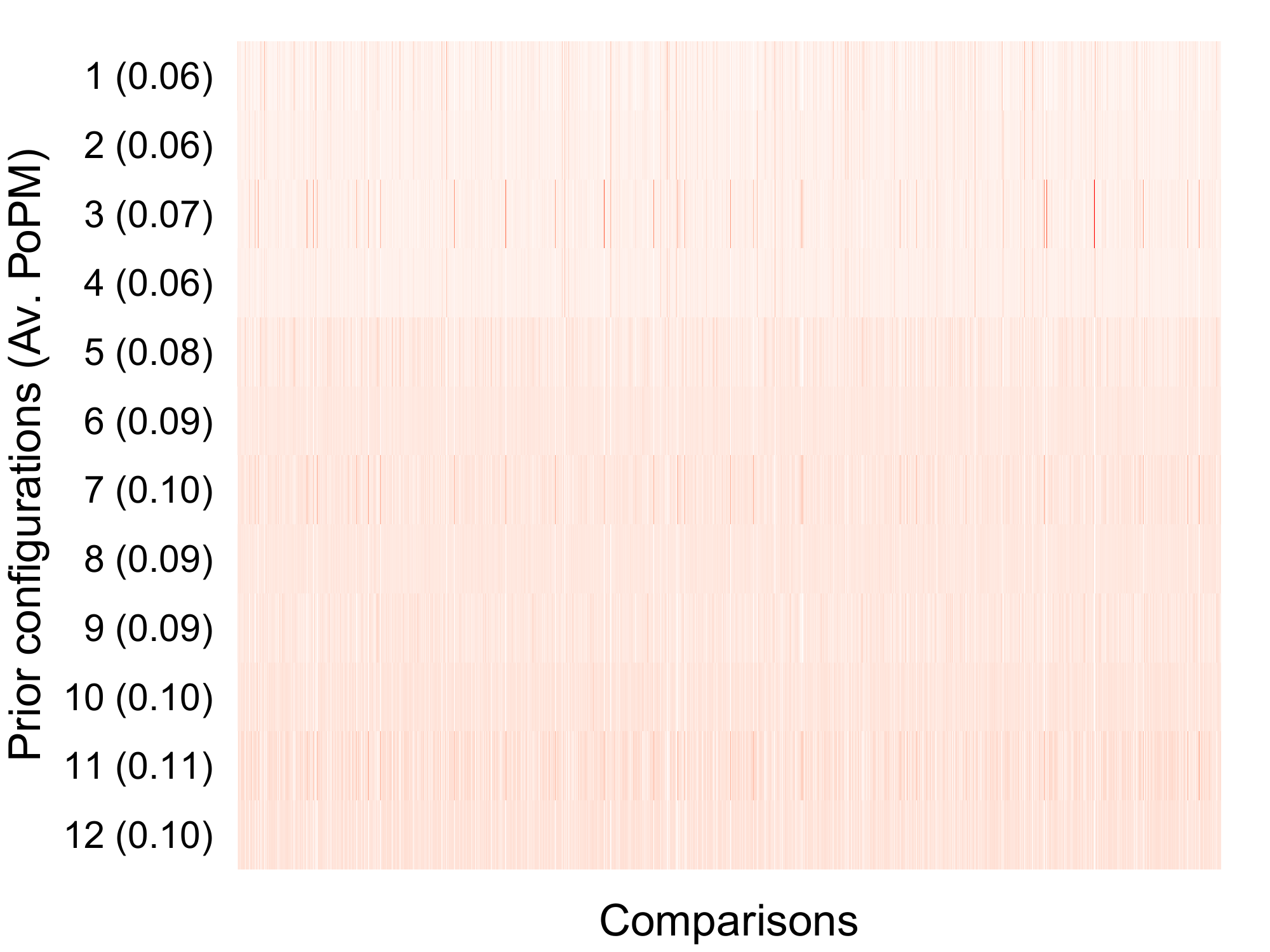}
\caption{Average posterior probabilities (AV. PoMP) for each of the 12 prior configurations under $\mathcal{H}_1^r$ for all 2406 test-set comparisons individually. For each comparison, the color gradient ranges from white (low posterior probability) to dark red (high posterior probability). The numbers in parentheses are the averaged posterior probabilities across all 2406 comparisons (conditional on $\mathcal{H}_1^r$). The prior probability for each configuration is $\nicefrac{1}{12} \approx 0.083$. See also Table~\ref{tab:ranking_prior_configurations}.}
\label{fig:ranking_prior_configurations}
\end{figure*}

Figure~\ref{fig:ranking_prior_configurations} displays the posterior probability for each of the 12 prior distribution configurations across the 2406 comparisons. The color gradient ranges from white (representing low posterior probability) to dark red (representing high posterior probability). Figure~\ref{fig:ranking_prior_configurations} shows that, on average, the different prior distribution configurations perform similarly. As suggested by the posterior rankings from Table~\ref{tab:ranking_prior_configurations}, configuration 1 predicted the data relatively poorly, resulting in an average posterior probability of $0.06$; in contrast, configuration 11 predicted the data relatively well, resulting in an average posterior probability of $0.11$. However, these posterior probabilities differ only modestly from the prior probability of $\nicefrac{1}{12} \approx 0.083$, and the Bayes factor associated with the comparison of posterior probabilities of $0.11$ and $0.06$ is less than 2. 

In sum, among the 12 prior configurations under $\mathcal{H}_1^r$ the best predictive performance was consistently obtained by data-driven priors, and the worst predictive performance was obtained by uninformed priors (cf. the posterior rankings from Table~\ref{tab:ranking_prior_configurations}). However, the extent of this predictive advantage is relatively modest: starting from a prior probability of $\nicefrac{1}{12} \approx 0.083$, 
the worst prior configuration has an average posterior probability of $0.06$, and the best prior configuration has average posterior probability of $0.11$ (cf. Figure~\ref{fig:ranking_prior_configurations}).\footnote{This modest change in plausibility is consistent with the fact that for overlapping prior distributions, there is an asymptotic limit to the evidence that is given by the ratio of prior ordinates evaluated at the maximum likelihood estimate.\cite{LyWagenmakerssubmPeri}}
 
\subsubsection{Posterior Probability of the Four Model Types}

In the second analysis we evaluate the predictive performance of the four meta-analytic model types (i.e., $\mathcal{H}_0^f$, $\mathcal{H}_1^f$, $\mathcal{H}_0^r$, and $\mathcal{H}_1^r$) by model-averaging across all prior distribution configurations, separately for each of the 2406 comparisons. Table~\ref{tab:prior_prob} shows the prior model probabilities obtained by first assigning probability $\nicefrac{1}{4}$ to each of the four model types, and then spreading that probability out evenly across the constituent prior distribution configurations.\cite[p. 47]{Jeffreys1961}

\begin{table*}[h]
    \centering
    \caption{Overview of the prior probability assignment to the different models and prior distribution configurations.}
    \label{tab:prior_prob}
    	\begin{tabular}{ccllc}
    		\toprule
    		 & Prior Model & & & Prior Configuration \\
    	            Model  &   Probability                        & Effect Size $\delta$ & Heterogeneity $\tau$ & Probability\\
    	               \midrule
    	               $\mathcal{H}_0^f$ & $\nicefrac{1}{4}$ & $\delta = 0$ & $\tau = 0$ & $\nicefrac{1}{4}$ \\
    	               \midrule
    	               \multirow{3}{*}{$\mathcal{H}_1^f$} & \multirow{3}{*}{$\nicefrac{1}{4}$} & $\delta \sim \text{Cauchy}(0, \nicefrac{1}{\sqrt{2}})$ & $\tau = 0$ & $\nicefrac{1}{12}$ \\
    	               &  & $\delta \sim \mathcal{N}(0, 0.56^2)$ & $\tau = 0$ & $\nicefrac{1}{12}$ \\
    	               &  & $\delta \sim \mathcal{T}(0, 0.33, 3)$ & $\tau = 0$ & $\nicefrac{1}{12}$ \\
    	               \midrule
    	               \multirow{4}{*}{$\mathcal{H}_0^r$} & \multirow{4}{*}{$\nicefrac{1}{4}$} & $\delta = 0$ & $\tau \sim \mathcal{U}(0, 1)$ & $\nicefrac{1}{16}$ \\
    	               & & $\delta = 0$ & $\tau \sim \mathcal{N}_+(0, 0.57^2)$ & $\nicefrac{1}{16}$ \\
    	               & & $\delta = 0$ & $\tau \sim \text{Inv-Gamma}(1.26, 0.24)$ & $\nicefrac{1}{16}$ \\
    	               & & $\delta = 0$ & $\tau \sim \text{Gamma}(1.59, 0.26)$ & $\nicefrac{1}{16}$ \\
    	               \midrule
    	               \multirow{12}{*}{$\mathcal{H}_1^r$} & \multirow{12}{*}{$\nicefrac{1}{4}$} & $\delta \sim \text{Cauchy}(0, \nicefrac{1}{\sqrt{2}})$ & $\tau \sim \mathcal{U}(0, 1)$ & $\nicefrac{1}{48}$ \\
    	               & & $\delta \sim \text{Cauchy}(0, \nicefrac{1}{\sqrt{2}})$ & $\tau \sim \mathcal{N}_+(0, 0.57^2)$ & $\nicefrac{1}{48}$ \\
    	               & & $\delta \sim \text{Cauchy}(0, \nicefrac{1}{\sqrt{2}})$ & $\tau \sim \text{Inv-Gamma}(1.26, 0.24)$ & $\nicefrac{1}{48}$ \\
    	               & & $\delta \sim \text{Cauchy}(0, \nicefrac{1}{\sqrt{2}})$ & $\tau \sim \text{Gamma}(1.59, 0.26)$ & $\nicefrac{1}{48}$ \\
    	               
    	               & & $\delta \sim \mathcal{N}(0, 0.56^2)$ & $\tau \sim \mathcal{U}(0, 1)$ & $\nicefrac{1}{48}$ \\
    	               & & $\delta \sim \mathcal{N}(0, 0.56^2)$ & $\tau \sim \mathcal{N}_+(0, 0.57^2)$ & $\nicefrac{1}{48}$ \\
    	               & & $\delta \sim \mathcal{N}(0, 0.56^2)$ & $\tau \sim \text{Inv-Gamma}(1.26, 0.24)$ & $\nicefrac{1}{48}$ \\
    	               & & $\delta \sim \mathcal{N}(0, 0.56^2)$ & $\tau \sim \text{Gamma}(1.59, 0.26)$ & $\nicefrac{1}{48}$ \\
    	               
    	               & & $\delta \sim \mathcal{T}(0, 0.33, 3)$ & $\tau \sim \mathcal{U}(0, 1)$ & $\nicefrac{1}{48}$ \\
    	               & & $\delta \sim \mathcal{T}(0, 0.33, 3)$ & $\tau \sim \mathcal{N}_+(0, 0.57^2)$ & $\nicefrac{1}{48}$ \\
    	               & & $\delta \sim \mathcal{T}(0, 0.33, 3)$ & $\tau \sim \text{Inv-Gamma}(1.26, 0.24)$ & $\nicefrac{1}{48}$ \\
    	               & & $\delta \sim \mathcal{T}(0, 0.33, 3)$ & $\tau \sim \text{Gamma}(1.59, 0.26)$ & $\nicefrac{1}{48}$ \\
    		\bottomrule
    	\end{tabular}
\end{table*}

For any particular comparison, a model type's model-averaged posterior probability is obtained by summing the posterior probabilities of the constituent prior distribution configurations. For example, the model-averaged posterior probability for $\mathcal{H}_1^r$ is obtained by summing the posterior probabilities for the 12 possible prior configurations, each of them associated with prior probability $\nicefrac{1}{48}$ (cf. Table~\ref{tab:prior_prob}).

Table~\ref{tab:ranking_models} lists the four model types and summarizes the number of times their model-averaged posterior probability was ranked $1,2,\ldots,4$. The results show that, across all comparisons, complex models generally received more support than simple models. The model that predicted the data best was $\mathcal{H}_1^r$, the random-effects model that assumes the presence of an effect; the model that predicted the data worst was $\mathcal{H}_0^f$, the fixed-effect model that assumes the absence of an effect. However, even $\mathcal{H}_0^f$ outpredicted the other three model types in $662/2406 \approx 28$\% of comparisons. Table~\ref{tab:ranking_models} also shows the model-averaged posterior model probability across all comparisons. In line with the ranking results, the average probability for $\mathcal{H}_0^f$ decreased from $0.25$ to $0.19$, whereas that for $\mathcal{H}_1^r$ increased from $0.25$ to $0.36$. Nevertheless, the support for $\mathcal{H}_1^r$ across all comparisons is not overwhelming and does not appear to provide an empirical license to ignore $\mathcal{H}_0^f$ (or any of the other three model types) from the outset.

\begin{table*}[h]
    \centering
    \caption{Ranking totals for each model type based on the 2406 comparisons in the test set. The numbers indicate how many times a specific model type attained a specific posterior probability rank. Rank `1' represents the best performance. The rankings reflect predictive adequacy that is model-averaged across the possible prior distribution configurations (cf. Table~\ref{tab:prior_prob}).}
    \label{tab:ranking_models}
    \begin{tabular}{@{}lllllll@{}}
        \toprule
                             & \multicolumn{4}{c}{Rank}  &       &            \\
        Model                &    1 &    2 &    3 &    4 & PrMP* & AV. PoMP** \\ \midrule
        $\mathcal{H}_0^f$    &  662 &  177 &  183 & 1382 & .25   & .19        \\
        $\mathcal{H}_1^f$    &  573 &  334 & 1235 &  262 & .25   & .22        \\
        $\mathcal{H}_0^r$    &  406 & 1158 &  790 &   52 & .25   & .24        \\
        $\mathcal{H}_1^r$    &  765 &  737 &  196 &  708 & .25   & .36        \\ \bottomrule
        \multicolumn{7}{l}{*Prior model probability} \\
        \multicolumn{7}{l}{**Average posterior model probability} \\
    \end{tabular}
\end{table*}

\begin{figure}[h]
\begin{tabular}{cc}
    \begin{minipage}{.50 \textwidth}
    \includegraphics[width=1\textwidth]{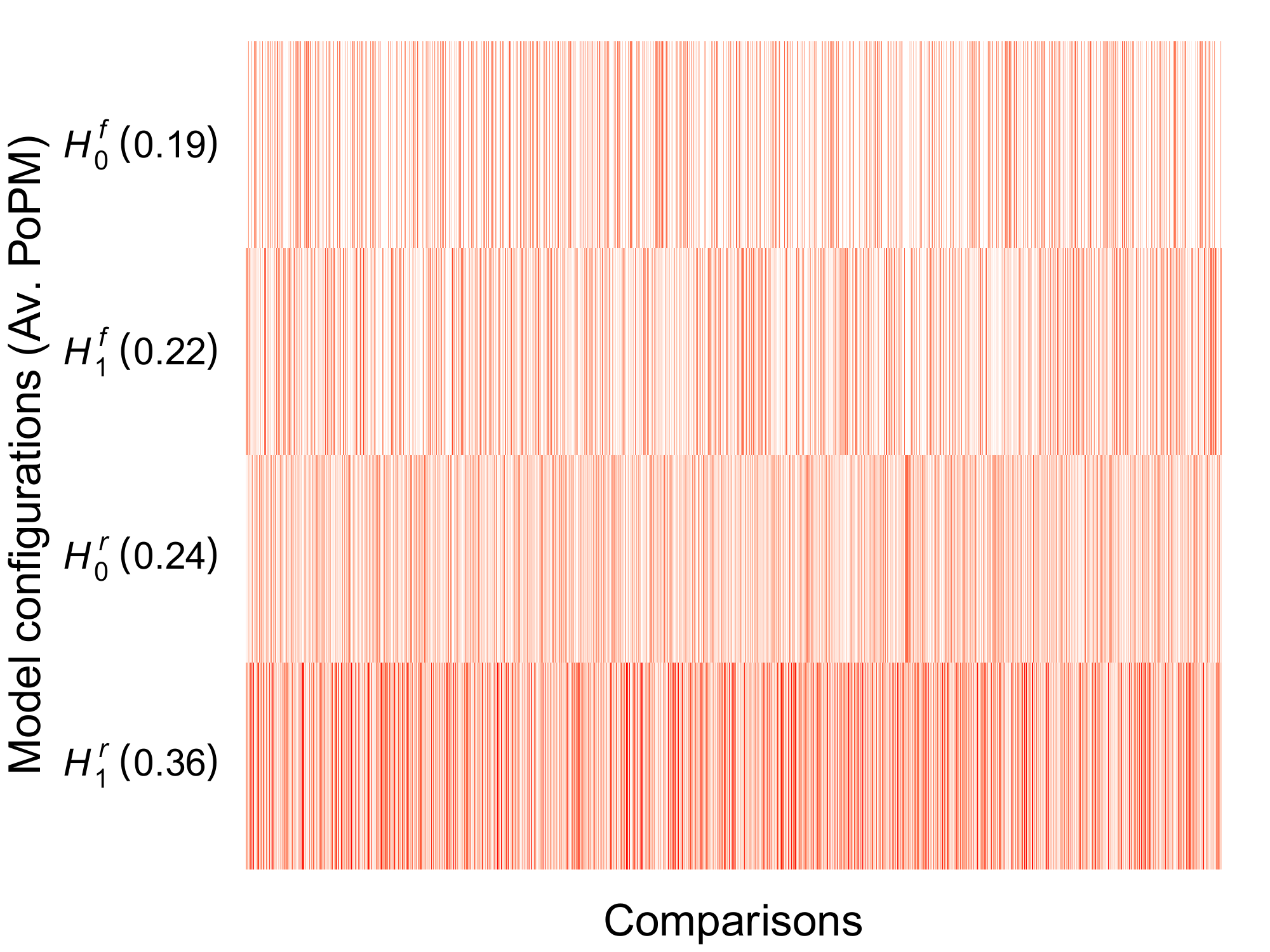}
    \end{minipage}  & %
    \begin{minipage}{.50 \textwidth}
    \includegraphics[width=1\textwidth]{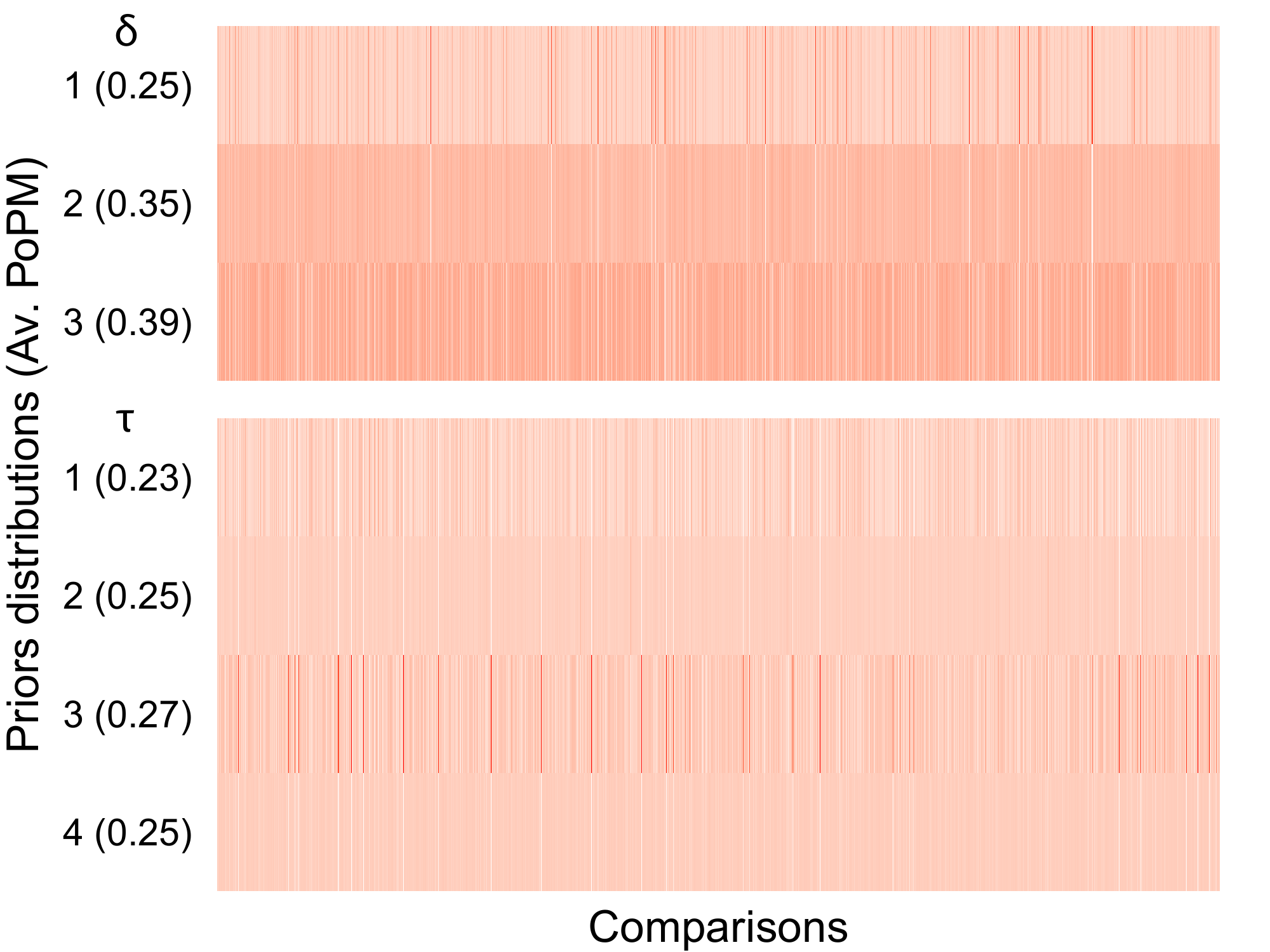}
    \end{minipage}  %
\end{tabular}
\caption{Model-averaged posterior probabilities (Av. PoMP) for each of the four model types for all 2406 test-set comparisons individually (left) and each prior distribution for all 2406 test-set comparisons individually (right). For each comparison, the color gradient ranges from white (low posterior probability) to dark red (high posterior probability). The numbers in parentheses are the averaged posterior probabilities across all 2406 comparisons. In the left panel, the prior probability for each model type is $\nicefrac{1}{4}$, see also Table~\ref{tab:ranking_models}. In the right panel, the prior probability is $\nicefrac{1}{3}$ for each prior distribution on $\delta$, and $\nicefrac{1}{4}$ for each prior distribution on $\tau$, see also Table~\ref{tab:ranking_priors}.}
\label{fig:heatmap_models_and_priors}
\end{figure}

Left panel of Figure~\ref{fig:heatmap_models_and_priors} displays the model-averaged posterior probability for each model type across the 2406 comparisons. It is apparent that the posterior probability is highest for  $\mathcal{H}_1^r$. However, for a substantial number of comparisons (i.e., $(662+573+406)/2406 \approx 68$\%, cf. Table~\ref{tab:ranking_models}) a different model type performs better. For instance --and in contrast to popular belief-- the fixed-effect models $\mathcal{H}_1^f$ and $\mathcal{H}_0^f$ together show the best predictive performance in a slight majority of comparisons (i.e., $(662+573)/2406 \approx 51$\%).

\subsubsection{Inclusion Bayes Factors}

\begin{figure}[h]
\begin{tabular}{cc}
    \begin{minipage}{.50 \textwidth}
    \includegraphics[width=1\textwidth]{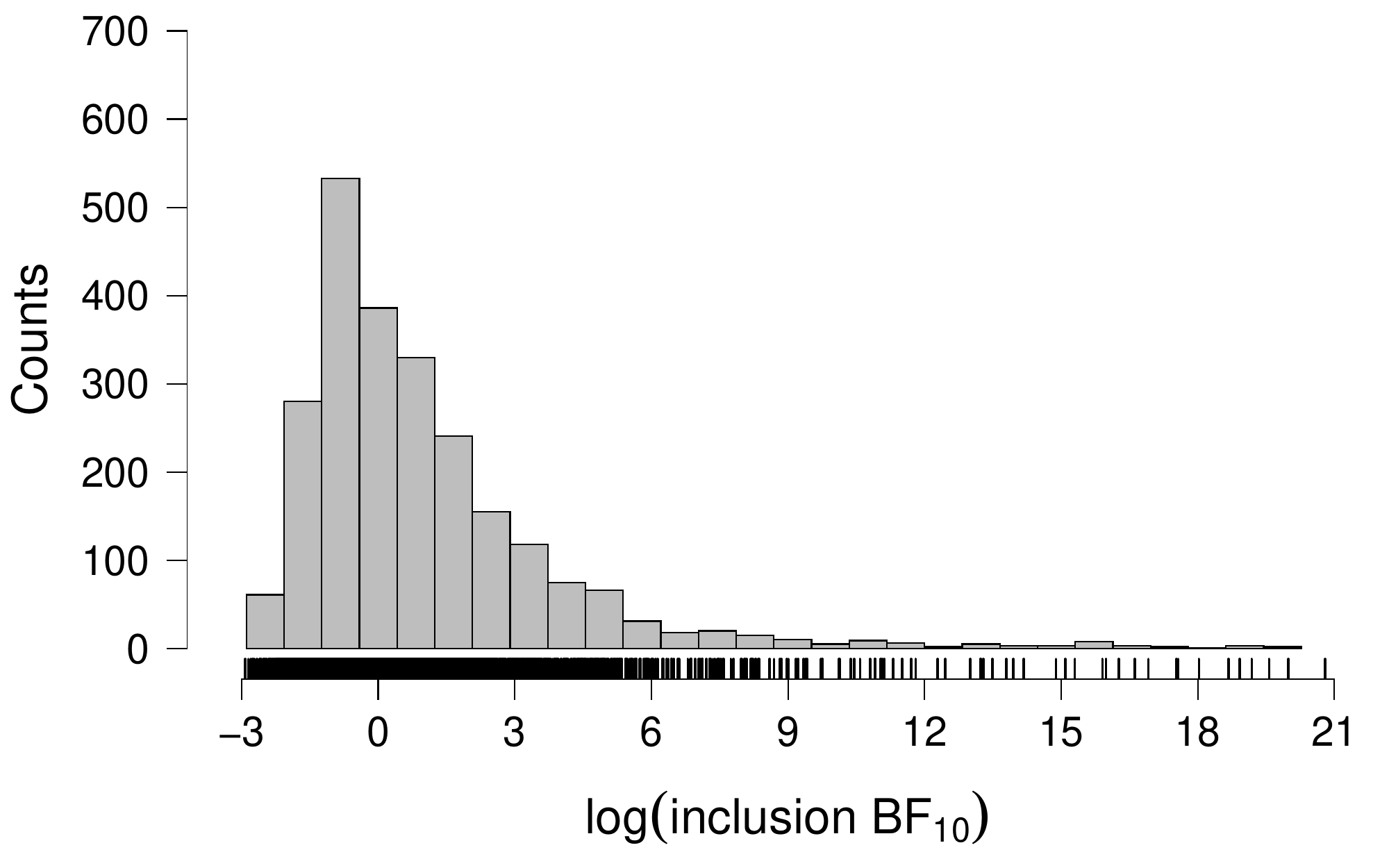}
    \end{minipage}  & %
    \begin{minipage}{.50 \textwidth}
    \includegraphics[width=1\textwidth]{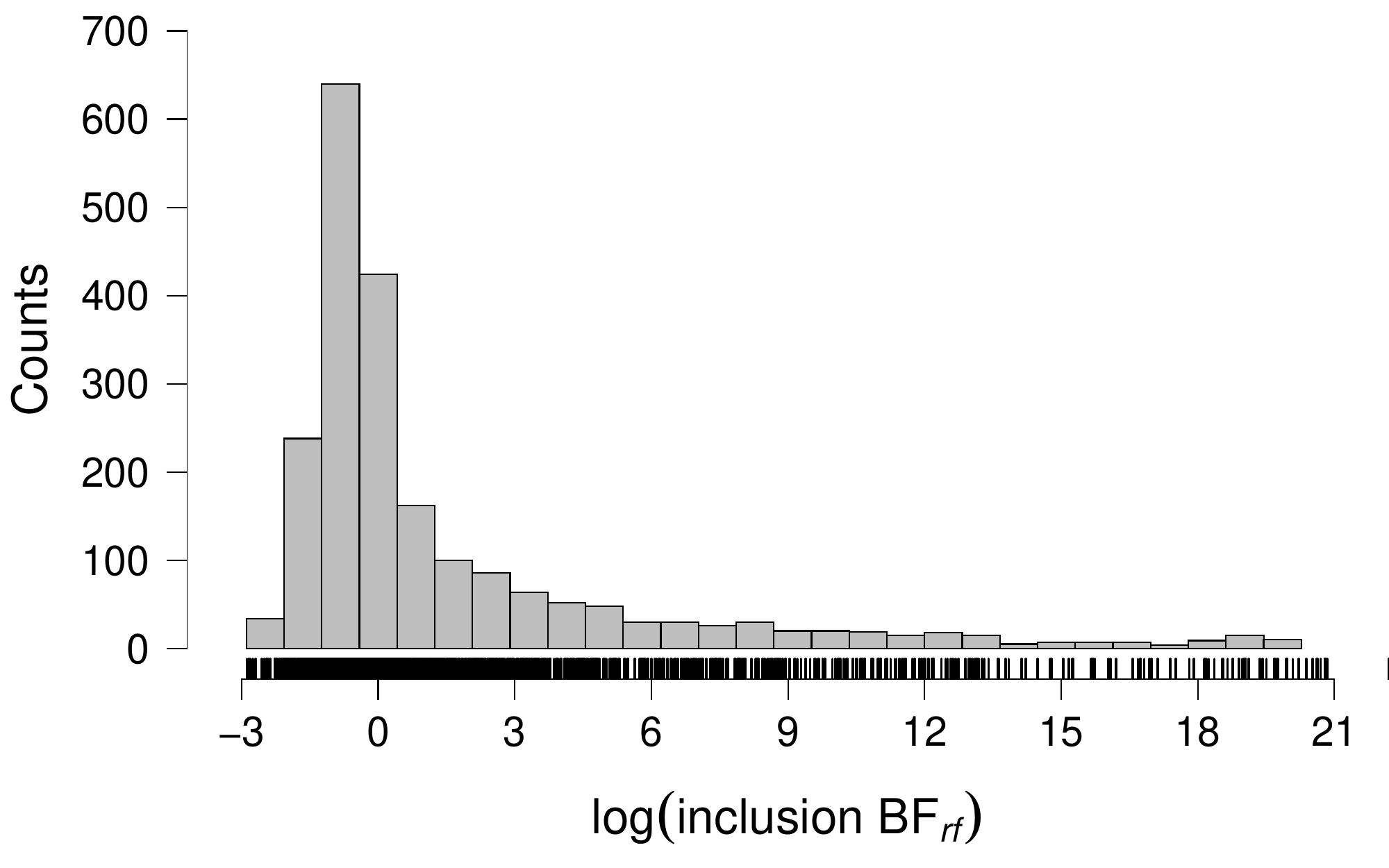}
    \end{minipage}  %
\end{tabular}
\caption{Inclusion Bayes factors in favor of the presence of a treatment effect (left) and in favor of the presence of across-study heterogeneity (right) for the 2406 comparisons in the test set. Not shown are log Bayes factors that exceed 21: twelve log Bayes factors for the presence of a treatment effect and 255 log Bayes factors for the presence of heterogeneity, or are lower than $-3$: one log Bayes factors for the presence of a treatment effect and four log Bayes factors for the presence of heterogeneity.}
\label{fig:inclusion-BF}
\end{figure}

In the third analysis we assess the inclusion Bayes factors for a treatment effect (cf. Eq.~\ref{eq:BF_treatment}) and for heterogeneity (cf. Eq.~\ref{eq:BF_heterogeneity}); that is, we model-average across all prior distribution configurations and across two model types, separately for each of the 2406 comparisons. First, the inclusion $\text{BF}_{10}$ for a treatment effect quantifies the evidence that the data provide for the presence vs. the absence of a group-level effect, taking into account the model uncertainty associated with whether the effect is fixed or random. The left panel of Figure~\ref{fig:inclusion-BF} displays a histogram of the log of the model-averaged $\text{BF}_{10}$ for the test set featuring 2406 comparisons. The histogram is noticeably right-skewed, which affirms the regularity  that it is easier to obtain compelling evidence for the presence rather than the absence of an effect.\cite[p. 196-197;]{Jeffreys1939}\footnote{This is the case because most prior distributions assign considerable mass to values near the test-relevant one; it is not a universal property of Bayes factors.\cite{JohnsonRossell2010}} Evidence for the presence of an effect was obtained in a small majority of the comparisons (i.e., $1336/2406 \approx 55.5\%$). 

Second, the inclusion $\text{BF}_{\text{rf}}$ for heterogeneity quantifies the evidence that the data provide for the presence vs. absence of between-study variability, taking into account the model uncertainty associated with whether the group-level effect is present or absent. The right panel of Figure~\ref{fig:inclusion-BF} displays a histogram of the log of the model-averaged $\text{BF}_{\text{rf}}$ for the test set featuring 2406 comparisons. The right-skew again confirms the regularity: it is easier to find compelling evidence for heterogeneity than for homogeneity. Nevertheless, the data provide evidence for heterogeneity only in a slight majority of $1227/2406 \approx 51.0\%$ of the comparisons.

In sum, the inclusion Bayes factors revealed that in nearly half of the comparisons from the test set, the data provide evidence in favor of the absence of a treatment effect (i.e., $44.5$\%) and provide evidence in favor of the absence of heterogeneity (i.e., $49.0$\%). The distribution of the log Bayes factors is asymmetric, indicating that it is easier to obtain compelling evidence for the presence of a treatment effect (rather than for its absence) and for the presence of heterogeneity (rather than for homogeneity).

\subsection{Exploratory Analysis: Model-Averaging Across Prior Distributions Under $\mathcal{H}_1^r$}

To further investigate the predictive performance of the prior distributions, we performed one additional analysis that was not preregistered in the original analysis plan. We focused on $\mathcal{H}_1^r$ and evaluated the prior distributions for each parameter by model-averaging across the possible prior distributions for the other parameter. For instance, to obtain the model-averaged posterior probability for the Cauchy prior distribution on the $\delta$ parameter (i.e., $\delta \sim \text{Cauchy}(0,\nicefrac{1}{\sqrt{2}})$), we consider the posterior probability for all 12 possible prior configurations and then sum across the four possible prior distributions on the $\tau$ parameter -- the four top models listed in the $\mathcal{H}_1^r$ row of Table~\ref{tab:prior_prob}. This way we obtain an assessment of the relative predictive performance of a particular prior distribution, averaging over the uncertainty on the prior distribution for the other parameter.

\begin{table*}[h]
    \centering
    \caption{Ranking totals for each prior distribution in $\mathcal{H}_1^r$ based on the 2406 comparisons in the test set. The numbers indicate how many times a specific prior distribution attained a specific posterior probability rank. Rank `1' represents the best performance. The rankings reflect predictive adequacy that is model-averaged across the possible prior distribution configurations of the other parameter.}
    \label{tab:ranking_priors}
    \begin{tabular}{@{}lllllll@{}}
        \toprule
                                                  & \multicolumn{4}{c}{Rank}  &       &            \\
        Prior distribution                        &    1 &    2 &    3 &    4 & PrMP* & AV. PoMP** \\
        \\
        \multicolumn{7}{l}{Parameter $\delta$}                                                        \\
        \midrule
        $\text{Cauchy}(0,\nicefrac{1}{\sqrt{2}})$ &  142 &  199 & 2065 &    $-$  &   .33 &       0.25 \\ 
        $\mathcal{N}(0, 0.56^2)$                  &  655 & 1727 &   24 &    $-$  &   .33 &       0.35 \\
        $\mathcal{T}(0, 0.33, 3)$  & 1609 &  480 &  317 &    $-$  &   .33 &       0.39 \\ 
        \\
        \multicolumn{7}{l}{Parameter $\tau$}                                                       \\
        \midrule
        $\mathcal{U}(0, 1)$                       &  576 &   83 &  116 & 1631 &   .25 &       0.23 \\ 
        $\mathcal{N}_+(0, 0.57^2)$                &   47 &  772 & 1587 &    0 &   .25 &       0.25 \\  
        $\text{Inv-Gamma}(1.26, 0.24)$            & 1418 &  172 &   67 &  749 &   .25 &       0.27 \\  
        $\text{Gamma}(1.59, 0.26)$                &  365 & 1379 &  636 &   26 &   .25 &       0.25 \\  
        \bottomrule
        \multicolumn{7}{l}{*Prior model probability} \\
        \multicolumn{7}{l}{**Average posterior model probability} \\
    \end{tabular}
\end{table*}

Table~\ref{tab:ranking_priors} lists the prior distributions and gives the number of times their model-averaged posterior probability attained a particular ranking. Consistent with the results reported earlier, the more data-driven prior distributions generally received more support than the prior distributions that are less informed. For the $\delta$ parameter, the best performing prior distribution was $\delta \sim \mathcal{T}(0, 0.33, 3)$; for the $\tau$ parameter, the best performing prior distribution was $\tau \sim \text{Inv-Gamma}(1.26, 0.24)$. Although the preference for the data-driven prior distributions is relatively consistent, it is not particularly pronounced, echoing the earlier results. Specifically, Table~\ref{tab:ranking_priors} also shows the model-averaged posterior model probability across all comparisons. For the $\delta$ parameter, the $t$-prior has a model-averaged posterior probability of $0.39$ (up from \nicefrac{1}{3}), but the Cauchy prior retains a non-negligible probability of $0.25$. For the $\tau$ parameter, the different prior distributions perform even more similarly; on average, the worst prior distribution is $\tau \sim \mathcal{U}(0, 1)$, and yet its model-averaged posterior model probability equals $0.23$, down from $\nicefrac{1}{4}$ but only a little. Likewise, the on-average best prior distribution is $\tau \sim \text{Inv-Gamma}(1.26, 0.24)$, with a model-averaged posterior model probability of $0.27$, which is only modestly larger than $\nicefrac{1}{4}$.

Right panel of Figure~\ref{fig:heatmap_models_and_priors} displays the model-averaged posterior probability for each prior distribution across the 2406 comparisons. The figure confirms that the data-driven prior distributions perform somewhat better than the relatively uninformed prior distributions. The color band is darker red, on average, for the prior distributions with the highest posterior model probabilities, that is, $\delta \sim \mathcal{T}(0, 0.33, 3)$ and $\tau \sim \text{Inv-Gamma}(1.26, 0.24)$.

\section{Exploratory Analysis: Subfield-Specific Prior Distributions}

Medical subfields may differ both in the typical size of the effects and in their degree of heterogeneity. In recognition of this fact we sought to develop empirical prior distributions for $\delta$ and $\tau$ that are subfield-specific. We differentiated between 47 medical subfields according to the taxonomy of the Cochrane Review Group. Based on their relatively good predictive performance detailed in the previous sections, we selected a $t$-distribution for the $\delta$ parameter (i.e., for subfield $i$, $\delta_i \sim \mathcal{T}(\mu=0, \sigma_i, \nu_i)$) and an inverse-gamma distribution for the $\tau$ parameter (i.e., for subfield $i$, $\tau_i \sim \text{Inverse-gamma}(\alpha_i, \beta_i)$).  

To estimate the parameters of these distributions separately for each subfield, we used the complete data set and proceeded analogously to the training set preparation: we removed comparisons with non-estimable studies, only used comparisons with at least ten studies, re-estimated the comparisons with a restricted maximum likelihood estimator, and removed comparisons with $\tau < 0.01$ estimates. These frequentist estimates were used as input for constructing the data-driven subfield-specific prior distributions. However, since many subfields contain only a limited number of comparisons, we used Bayesian hierarchical estimation with weakly informative priors on the hyperparameters. The hierarchical aspect of the estimation procedure shrinks the estimated parameter values towards the grand mean, a tendency that is more pronounced if the estimated field-specific value is both extreme and based on relatively little information.\cite{LeeWagenmakersBayesBook,McElreath2016,RouderLu2005} Specifically, we assumed that all field-specific parameters (i.e., $\sigma_i$, $\nu_i$, $\alpha_i$, and $\beta_i$) are governed by positive-only normal distributions. For the $t$-distribution, we assigned positive-only Cauchy$(0,k)$ prior distributions both to the across-field normal mean and to the across-field normal standard deviation, with $k=1$ for parameter $\sigma$ and $k=10$ for parameter $\nu$. For the inverse-gamma distribution, we assigned positive-only Cauchy$(0,1)$ prior distributions both to the across-field normal mean and to the across-field normal standard deviation for shape parameter $\alpha$ and scale parameter $\beta$. The hierarchical models were estimated using the \texttt{rstan} \texttt{R} package\cite{rstan} that interfaces with the Stan probabilistic modeling language.\cite{stan} The Stan code is available alongside the supplementary materials at \url{https://osf.io/zs3df/}.

\begin{figure}[h]
\begin{tabular}{cc}
    \begin{minipage}{.50 \textwidth}
    \includegraphics[width=1\textwidth]{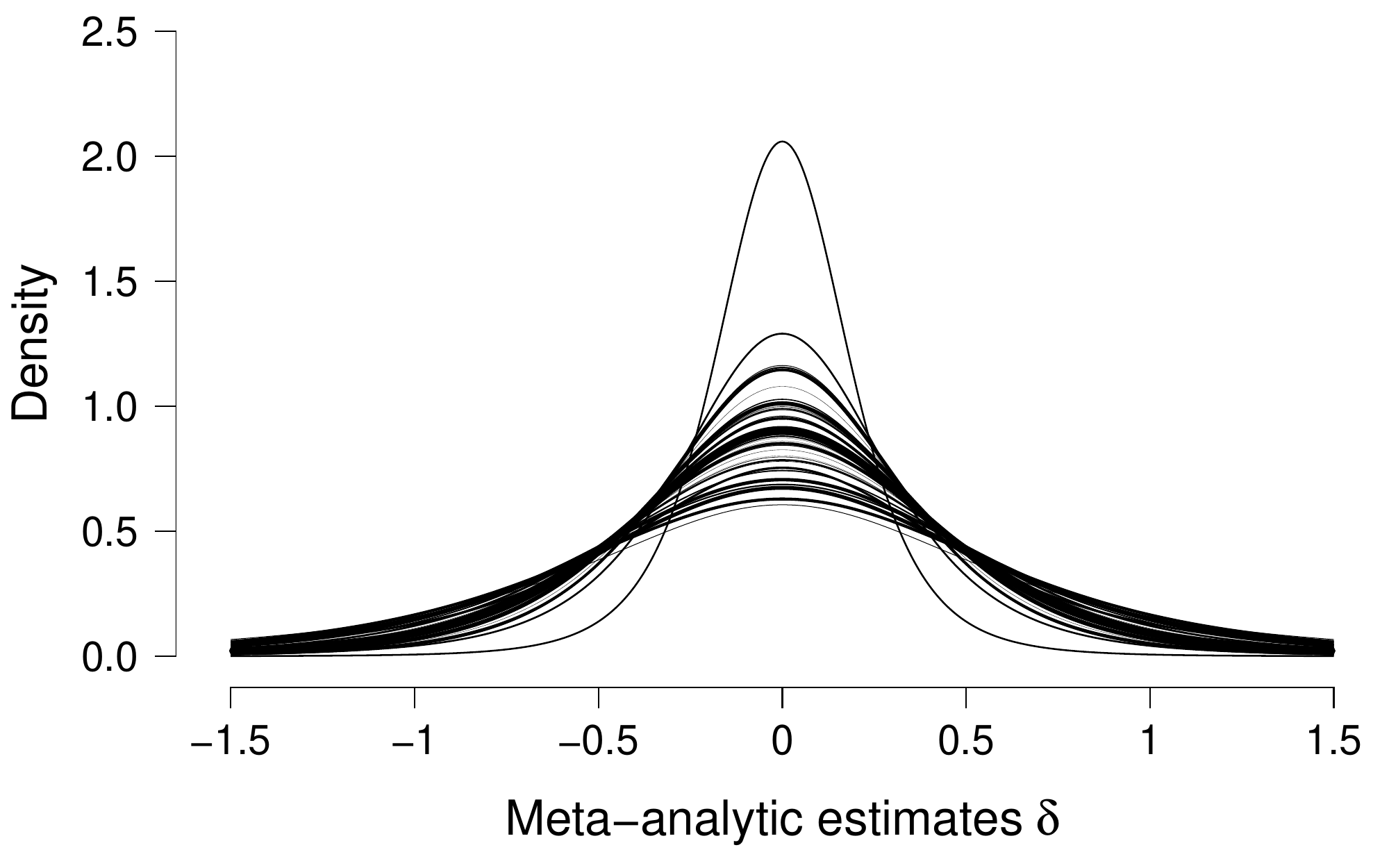}
    \end{minipage}  & %
    \begin{minipage}{.50 \textwidth}
    \includegraphics[width=1\textwidth]{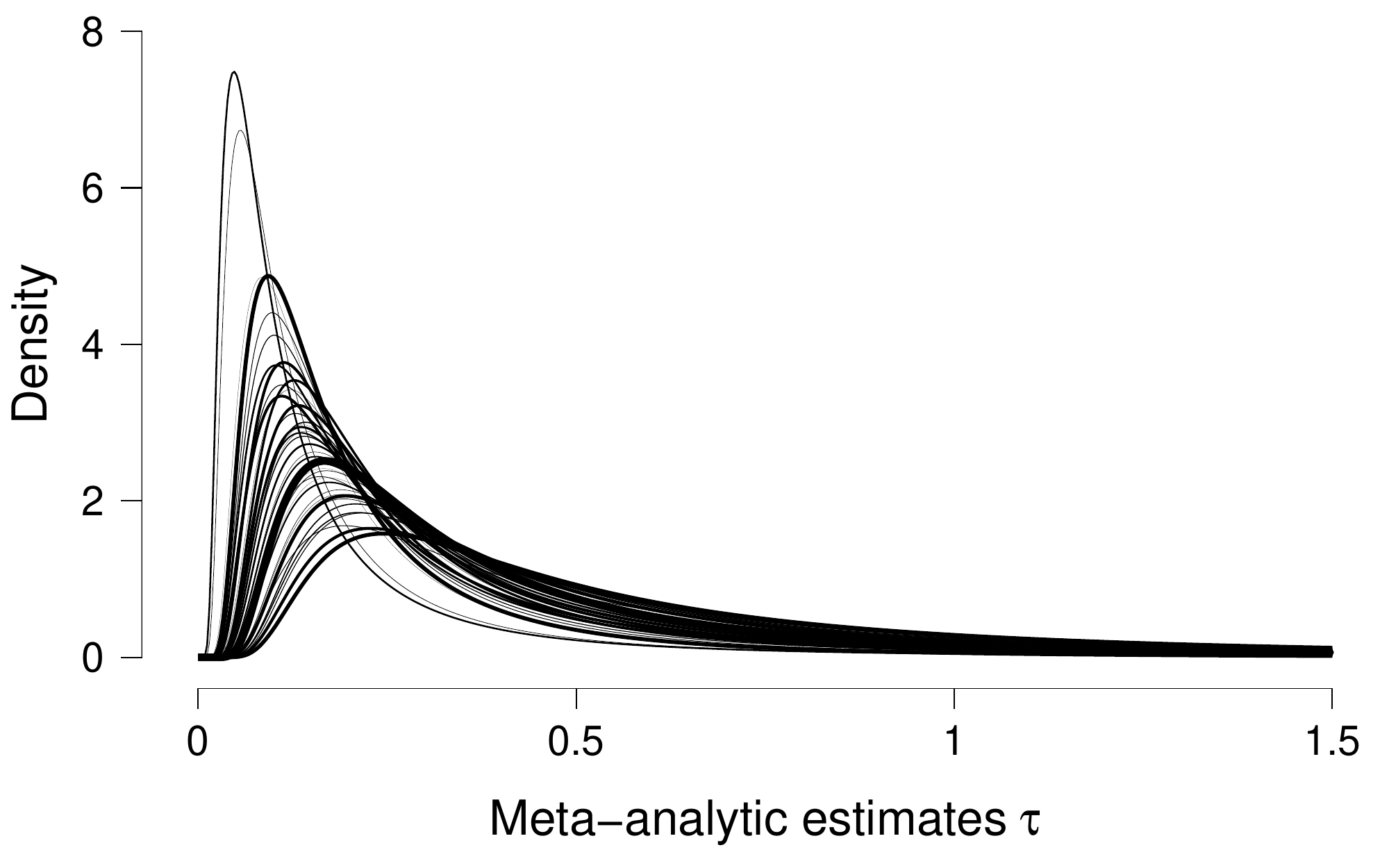}
    \end{minipage}  %
\end{tabular}
\caption{Subfield-specific prior distributions for parameter $\delta$ (left panel) and parameter $\tau$ (right panel) for 46 individual topics from the Cochrane database of systematic reviews estimated by hierarchical regression based on the complete data set. See also Table~\ref{tab:differences_fields_priors}.}
\label{fig:subfields}
\end{figure}

Table~\ref{tab:differences_fields_priors} lists the 46 different subfields (the 47\textsuperscript{th} subfield ``Multiple Sclerosis and Rare Diseases of the CNS'' featured two comparisons, both of which were excluded based on the $\tau < 0.01$ criterion), the associated number of comparisons and studies, and the estimated distributions for both $\delta$ and $\tau$. The scale estimates for the $\delta$ parameter show considerable variation, ranging from $0.18$ (``Developmental, Psychosocial and Learning Problems'') to 0.60 (``Hepato-Biliary'').\footnote{Notice that the most extreme estimates come from fields with relatively large number of comparisons, as these estimates are less subject to shrinkage.} A similar variation is present in the estimated distributions for the $\tau$ parameter. Figure~\ref{fig:subfields} visualizes the prior distributions for each subfield.

\begin{table}[ph!]
\centering
\caption{Subfield-specific prior distributions for 46 individual topics from the Cochrane database of systematic reviews estimated by hierarchical regression based on the complete data set. The $t$-distribution follows a location, scale, and degrees of freedom parametrization and the inverse-gamma distribution follows a shape and scale parametrization. See also Figure~\ref{fig:subfields}.}
\label{tab:differences_fields_priors}
\begin{tabular}{lrrrr}
Topic & Comparisons & Studies & Prior $\delta$ & Prior $\tau$ \\
\toprule
Acute Respiratory Infections & 6 & 104 & $\mathcal{T}(0, 0.38, 5)$ & $\text{Inv-Gamma}(1.73, 0.46)$  \\  
Airways & 46 & 815 & $\mathcal{T}(0, 0.38, 6)$ & $\text{Inv-Gamma}(2.02, 0.28)$  \\  
Anaesthesia & 44 & 661 & $\mathcal{T}(0, 0.55, 4)$ & $\text{Inv-Gamma}(1.62, 0.64)$  \\  
Back and Neck & 13 & 278 & $\mathcal{T}(0, 0.37, 5)$ & $\text{Inv-Gamma}(1.75, 0.57)$  \\  
Bone, Joint and Muscle Trauma & 32 & 1221 & $\mathcal{T}(0, 0.40, 5)$ & $\text{Inv-Gamma}(1.52, 0.28)$  \\  
Colorectal & 13 & 372 & $\mathcal{T}(0, 0.51, 5)$ & $\text{Inv-Gamma}(1.64, 0.56)$  \\  
Common Mental Disorders & 17 & 264 & $\mathcal{T}(0, 0.55, 5)$ & $\text{Inv-Gamma}(1.62, 0.45)$  \\  
Consumers and Communication & 6 & 72 & $\mathcal{T}(0, 0.40, 5)$ & $\text{Inv-Gamma}(1.56, 0.14)$  \\  
Cystic Fibrosis and Genetic Disorders & 1 & 12 & $\mathcal{T}(0, 0.47, 5)$ & $\text{Inv-Gamma}(1.70, 0.45)$  \\  
Dementia and Cognitive Improvement & 9 & 197 & $\mathcal{T}(0, 0.45, 5)$ & $\text{Inv-Gamma}(1.71, 0.44)$  \\  
Developmental, Psychosocial and Learning Problems & 20 & 407 & $\mathcal{T}(0, 0.18, 5)$ & $\text{Inv-Gamma}(1.43, 0.12)$  \\  
Drugs and Alcohol & 8 & 170 & $\mathcal{T}(0, 0.33, 5)$ & $\text{Inv-Gamma}(1.89, 0.28)$  \\  
Effective Practice and Organisation of Care & 10 & 204 & $\mathcal{T}(0, 0.39, 5)$ & $\text{Inv-Gamma}(1.71, 0.35)$  \\  
Emergency and Critical Care & 9 & 214 & $\mathcal{T}(0, 0.39, 5)$ & $\text{Inv-Gamma}(1.62, 0.29)$  \\  
ENT & 17 & 273 & $\mathcal{T}(0, 0.43, 5)$ & $\text{Inv-Gamma}(1.85, 0.48)$  \\  
Eyes and Vision & 14 & 347 & $\mathcal{T}(0, 0.40, 6)$ & $\text{Inv-Gamma}(1.86, 0.41)$  \\  
Gynaecological, Neuro-oncology and Orphan Cancer & 1 & 10 & $\mathcal{T}(0, 0.45, 5)$ & $\text{Inv-Gamma}(1.67, 0.46)$  \\  
Gynaecology and Fertility & 14 & 253 & $\mathcal{T}(0, 0.38, 5)$ & $\text{Inv-Gamma}(1.78, 0.46)$  \\  
Heart & 88 & 2112 & $\mathcal{T}(0, 0.42, 5)$ & $\text{Inv-Gamma}(1.83, 0.47)$  \\  
Hepato-Biliary & 34 & 1103 & $\mathcal{T}(0, 0.60, 4)$ & $\text{Inv-Gamma}(1.56, 0.58)$  \\  
HIV/AIDS & 2 & 23 & $\mathcal{T}(0, 0.43, 5)$ & $\text{Inv-Gamma}(1.73, 0.44)$  \\  
Hypertension & 27 & 524 & $\mathcal{T}(0, 0.48, 3)$ & $\text{Inv-Gamma}(2.01, 0.38)$  \\  
Incontinence & 17 & 219 & $\mathcal{T}(0, 0.33, 6)$ & $\text{Inv-Gamma}(1.64, 0.36)$  \\  
Infectious Diseases & 8 & 150 & $\mathcal{T}(0, 0.59, 2)$ & $\text{Inv-Gamma}(1.28, 0.44)$  \\  
Inflammatory Bowel Disease & 1 & 12 & $\mathcal{T}(0, 0.40, 5)$ & $\text{Inv-Gamma}(1.76, 0.39)$  \\  
Injuries & 3 & 54 & $\mathcal{T}(0, 0.35, 5)$ & $\text{Inv-Gamma}(1.80, 0.34)$  \\  
Kidney and Transplant & 39 & 767 & $\mathcal{T}(0, 0.54, 5)$ & $\text{Inv-Gamma}(1.72, 0.53)$  \\  
Metabolic and Endocrine Disorders & 25 & 503 & $\mathcal{T}(0, 0.43, 5)$ & $\text{Inv-Gamma}(1.71, 0.37)$  \\  
Methodology & 5 & 106 & $\mathcal{T}(0, 0.49, 5)$ & $\text{Inv-Gamma}(1.72, 0.51)$  \\  
Movement Disorders & 5 & 70 & $\mathcal{T}(0, 0.42, 5)$ & $\text{Inv-Gamma}(1.88, 0.33)$  \\  
Musculoskeletal & 32 & 778 & $\mathcal{T}(0, 0.45, 6)$ & $\text{Inv-Gamma}(1.87, 0.38)$  \\  
Neonatal & 11 & 259 & $\mathcal{T}(0, 0.42, 5)$ & $\text{Inv-Gamma}(1.68, 0.38)$  \\  
Oral Health & 10 & 236 & $\mathcal{T}(0, 0.51, 5)$ & $\text{Inv-Gamma}(1.79, 0.28)$  \\  
Pain, Palliative and Supportive Care & 16 & 283 & $\mathcal{T}(0, 0.43, 5)$ & $\text{Inv-Gamma}(1.69, 0.42)$  \\  
Pregnancy and Childbirth & 32 & 539 & $\mathcal{T}(0, 0.33, 5)$ & $\text{Inv-Gamma}(1.86, 0.32)$  \\  
Public Health & 2 & 22 & $\mathcal{T}(0, 0.33, 5)$ & $\text{Inv-Gamma}(1.76, 0.23)$  \\  
Schizophrenia & 21 & 436 & $\mathcal{T}(0, 0.29, 4)$ & $\text{Inv-Gamma}(1.60, 0.27)$  \\  
Sexually Transmitted Infections & 9 & 113 & $\mathcal{T}(0, 0.42, 5)$ & $\text{Inv-Gamma}(1.70, 0.59)$  \\  
Skin & 6 & 85 & $\mathcal{T}(0, 0.48, 5)$ & $\text{Inv-Gamma}(1.64, 0.51)$  \\  
Stroke & 21 & 357 & $\mathcal{T}(0, 0.48, 5)$ & $\text{Inv-Gamma}(1.71, 0.40)$  \\  
Tobacco Addiction & 4 & 44 & $\mathcal{T}(0, 0.44, 4)$ & $\text{Inv-Gamma}(1.73, 0.42)$  \\  
Upper GI and Pancreatic Diseases & 1 & 12 & $\mathcal{T}(0, 0.45, 5)$ & $\text{Inv-Gamma}(1.76, 0.38)$  \\  
Urology & 2 & 33 & $\mathcal{T}(0, 0.44, 5)$ & $\text{Inv-Gamma}(1.73, 0.45)$  \\  
Vascular & 3 & 35 & $\mathcal{T}(0, 0.46, 5)$ & $\text{Inv-Gamma}(1.66, 0.50)$  \\  
Work & 2 & 24 & $\mathcal{T}(0, 0.42, 5)$ & $\text{Inv-Gamma}(1.76, 0.39)$  \\  
Wounds & 7 & 103 & $\mathcal{T}(0, 0.56, 5)$ & $\text{Inv-Gamma}(1.54, 0.41)$  \\  
\midrule
Pooled estimate & 713 & 14876 & $\mathcal{T}(0, 0.43, 5)$ & $\text{Inv-Gamma}(1.71, 0.40)$  \\  
\bottomrule
\end{tabular}
\end{table}

\section{Example: Dentine Hypersensitivity}

We demonstrate BMA meta-analysis with an example from oral health. Poulsen et al.\cite{poulsen2006potassium} considered the effect of potassium-containing toothpaste on dentine hypersensitivity. Five studies with a tactile outcome assessment were subjected to a meta-analysis. In their review, Poulsen et al.\cite{poulsen2006potassium} reported a meta-analytic effect size estimate $\delta = 1.19$, 95\% CI $[0.79, 1.59]$, $z = 5.86$, $p < 0.00001$ of potassium-containing toothpastes on reducing tactile sensitivity (``Analysis 1.1. Comparison 1 Potassium containing toothpaste (update), Outcome 1 Tactile.'').

\begin{figure*}[h]
\centering
    \includegraphics[width=1\textwidth]{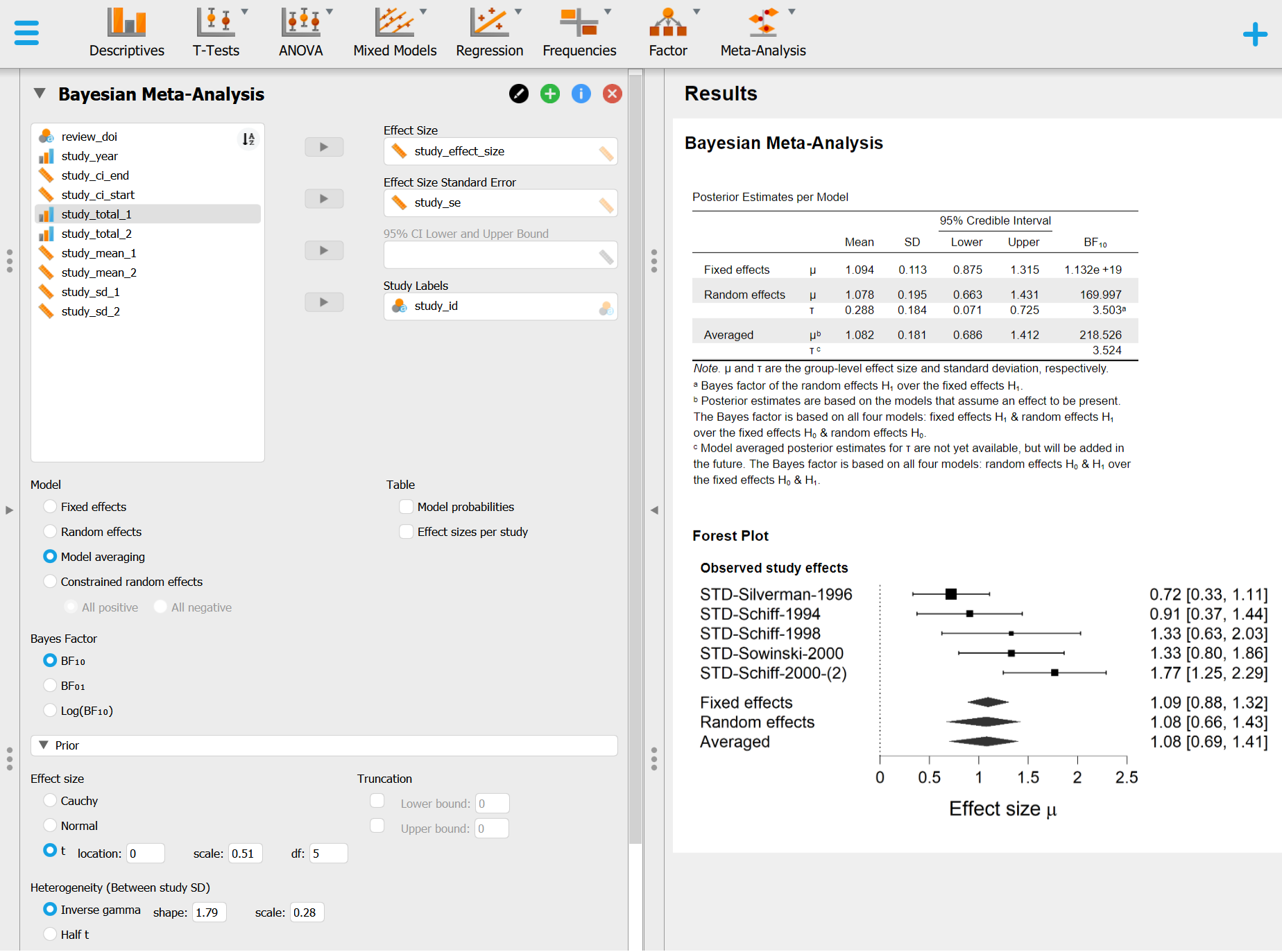}
    \caption{JASP screenshot of a Bayesian model-averaged meta-analysis of the Poulsen et al.\cite{poulsen2006potassium} comparison concerning the effect of potassium-containing toothpaste on dentine tactile hypersensitivity. The left input panel shows the specification of the ``Oral Health'' CDSR subfield-specific prior distributions for effect size $\delta$ and heterogeneity $\tau$. The right output panel shows the corresponding results.}
    \label{fig:JASP_example}
\end{figure*}

We re-analyze the Poulsen et al.\cite{poulsen2006potassium} comparison using the BMA meta-analysis implementation in the open-source statistical software package JASP (\url{jasp-stats.org}).\cite{JASP15,LoveEtAl2019JASP,LyEtAl2021ISBA,vanDoornEtAlinpressJASPGuidelines,WagenmakersEtAl2018PBRPartII} Appendix~\ref{app:example_R} provides the same analysis in \texttt{R}\cite{R} using the \texttt{metaBMA} package.\cite{heck2017metabma} Figure~\ref{fig:JASP_example} shows the JASP graphical user interface with the left panel specifying the analysis setting and the right panel displaying the default output. After loading the data into JASP, the BMA meta-analysis can be performed by activating the ``Meta-Analysis'' module after clicking  the blue ``+'' button in the top right corner, choosing ``Meta-Analysis'' from the ribbon at the top, and then selecting ``Bayesian Meta-Analysis`` from the drop-down menu. In the left input panel, we move the study effect sizes and standard errors into the appropriate boxes and adjust the prior distributions under the ``Prior'' tab to match the subfield-specific distributions given in Table~\ref{tab:differences_fields_priors}. Specifically, for the ``Oral Health'' subfield the prior distributions are $\delta \sim \mathcal{T}(0, 0.51, 5)$ and $\tau \sim \text{Inv-Gamma}(1.79, 0.28)$.  

The JASP output panel displays the corresponding BMA meta-analysis results. The ``Posterior Estimates per Model'' table summarizes the estimates and evidence from the fixed-effect models, random-effects models, and finally the model-averaged results. The final row of the table shows an effect size estimate $\delta = 1.082$, 95 \% CI $[0.686, 1.412]$ which is slightly lower and more conservative than the one provided by the frequentist random-effects meta-analysis, further quantified with extreme evidence for the presence of an effect, $\text{BF}_{10} = 218.53$, and moderate evidence for the presence of heterogeneity, $\text{BF}_{\text{rf}} = 3.52$. The JASP output panel also presents a forest plot that visualizes the observed effects size estimates from the individual studies, the overall fixed-effect and random-effects meta-analytic estimates and the corresponding model-averaged effect size estimate.

The JASP interface provides additional options not discussed here, such as (a) visualizing the prior and posterior distributions; (b) visualizing the estimated effect sizes from individual studies; (c) performing one-sided hypothesis tests; (d) updating evidence sequentially, study-by-study; (e) adding ordinal constraints\cite{haaf2020does}; and (f) adjustments for publication bias.\cite{maier2020robust, bartos2020adjusting}

\section{Concluding Comments}

In this article, we introduced Bayesian model-averaged meta-analysis for continuous outcomes in medicine. The proposed methodology provides a principled way to integrate, quantify, and update uncertainty regarding both parameters and models. Specifically, the methodology allows researchers to simultaneously test for and estimate effect size and heterogeneity without committing to a particular model in an all-or-none fashion. In Bayesian model-averaged meta-analysis, multiple models are considered simultaneously, and inference is proportioned to the support that each model receives from the data. This eliminates the need for stage-wise, multi-step inference procedures that first identify a single preferred model (e.g., a fixed-effect model or a random-effects model) and then interpret the model parameters without acknowledging the uncertainty inherent in the model selection stage. The multi-model approach advocated here also decreases the potential impact of model misspecification.

Bayesian model-averaged meta-analysis comes with the usual advantages of Bayesian statistics -- the ability to quantify evidence in favor or against any hypothesis (including the null hypothesis), the ability to discriminate between absence of evidence and evidence of absence,\cite{KeysersEtAl2020, Robinson2019} the ability to monitor evidence as individual studies accumulate,\cite{ter_schure_accumulation_2019} the straightforward interpretation of the results (i.e., probability statements that refer directly to parameters and hypotheses),\cite{OHaganForster2004} and the opportunity to incorporate historical information.\cite{berry2006bayesian, hobbs2007practical} In this article, our goal was to take advantage of the existing medical knowledge base in order to propose and assess prior distributions that allow for more efficient inference.

Following a preregistered analysis plan, we fitted and assessed different prior distributions for both effect size $\delta$ and heterogeneity $\tau$ using comparisons of continuous outcomes from the Cochrane database of systematic reviews. We fitted prior distributions based on a training set of randomly selected comparisons, and then evaluated predictive performance based on a test set. The results showed that predictive performance on the test set was relatively similar for the different data-driven prior distributions. Moreover, and in contrast to popular belief and recommendations,\cite{higgins2009re,chung2013avoiding,council1992combining, mosteller1996understanding} we did not find that the random-effects meta-analytic model provided a superior account of the data: the random-effects meta-analytic models outpredicted their fixed-effect counterparts in only 51.0\% of comparisons. Although the random-effects alternative hypothesis $\mathcal{H}_1^r$ showed the best predictive performance on average, the data increased its model-averaged posterior probability from $0.25$ to only $0.36$, leaving $0.64$ for the three competing model types (i.e., a model with no heterogeneity, a model without an effect, and a model without both).

Based on the outcome of our preregistered analysis, we used the data from Cochrane database of systematic reviews to develop empirical prior distributions for continuous outcomes in 46 different medical subfields. Finally, we applied Bayesian model-averaged meta-analysis with subfield-specific prior distributions to an example from oral health, using the free statistical software packages \texttt{R} and JASP. We believe that the proposed Bayesian methodology provides an alternative perspective on meta-analysis that is informed, efficient, and insightful.

\section*{Acknowledgments}
This work was supported by The Netherlands Organisation for Scientific Research (NWO) through a Research Talent grant (to QFG; 406.16.528), a Vici grant (to EJW; 016.Vici.170.083), and a NWA Idea Generator grant (to WMO; NWA.1228.191.045).


\section*{Financial disclosure}

None reported.

\section*{Data availability statement}

Data and analysis scripts are publicly available at: https://osf.io/zs3df/files/.

\section*{Conflict of interest}

František Bartoš, Alexander Ly, and Eric-Jan Wagenmakers declare their involvement in the open-source software package JASP (\url{https://jasp-stats.org}), a non-commercial, publicly-funded effort to make Bayesian statistics accessible to a broader group of
researchers and students.

\bibliographystyle{WileyNJD-AMA}
\bibliography{References.bib}

\appendix

\section{R Code for the Oral Health Example}
\label{app:example_R}

The result for the example analysis can also be obtained with the statistical programming language \texttt{R}.\cite{R} After initializing the \texttt{R} session, the \texttt{metaBMA} \texttt{R} package\cite{heck2017metabma} needs to be installed with the following command (this command needs to be executed only if the package is not already present):

\begin{verbatim}install.packages("metaBMA")\end{verbatim}

\noindent After the \texttt{metaBMA} has been installed, it must then be loaded into the session, and the data set (containing the effect sizes and standard errors from the individual studies) from the ``Tactile Sensitivity.csv'' file can be made available, as follows:

\begin{verbatim}
library("metaBMA")
data <- read.csv("Tactile Sensitivity.csv")
\end{verbatim}

\noindent In order to perform the BMA meta-analysis with the subfield-specific prior distributions, we first assign the column containing the effect sizes \texttt{study\_effect\_size} to the \texttt{y} argument, and the column containing standard errors \texttt{study\_effect\_size} to the \texttt{SE} argument (the \texttt{data = data} arguments specifies that both of the columns are located in the already loaded data set called \texttt{data}). Then, we specify the prior distributions for the effect size, the \texttt{d} argument, and the heterogeneity, the \texttt{tau} argument, accordingly to the ``Oral Health'' row in Table~\ref{tab:differences_fields_priors}. Finally, the \texttt{control} argument specifies an additional control argument for the Markov chain Monte Carlo (MCMC) sampler, increasing the target acceptance probability from the default value of $0.80$, often required in cases with a small number of studies (see \url{https://mc-stan.org/misc/warnings.html} for more information about Stan's warnings and possible solutions):  
 
\begin{verbatim}
fit_example <- meta_bma(y = study_effect_size, SE = study_se, data = data,
         d   = prior("t",        c(location = 0, scale = 0.51, nu = 5)),
         tau = prior("invgamma", c(shape = 1.79, scale = 0.28)),
         control = list(adapt_delta = .90))
\end{verbatim}

\noindent To obtain the numerical summaries of the estimated model, we just execute the name of the object containing the fitted model:

\begin{verbatim}
fit_example
\end{verbatim}

\noindent The resulting output corresponds to that given by JASP output (up to MCMC error):
\begin{verbatim}
> fit_example
### Meta-Analysis with Bayesian Model Averaging ###
   Fixed H0:  d = 0 
   Fixed H1:  d ~ 't' (location=0, scale=0.51, nu=5) with support on the interval [-Inf,Inf]. 
   Random H0: d   = 0,   
              tau ~  'invgamma' (shape=1.79, scale=0.28) with support on the interval [0,Inf]. 
   Random H1: d   ~ 't' (location=0, scale=0.51, nu=5) with support on the interval [-Inf,Inf]. 
              tau ~ 'invgamma' (shape=1.79, scale=0.28) with support on the interval [0,Inf]. 

# Bayes factors:
           (denominator)
(numerator) fixed_H0 fixed_H1 random_H0 random_H1
  fixed_H0  1.00e+00 8.83e-20  4.29e-18  2.52e-20
  fixed_H1  1.13e+19 1.00e+00  4.85e+01  2.85e-01
  random_H0 2.33e+17 2.06e-02  1.00e+00  5.88e-03
  random_H1 3.97e+19 3.50e+00  1.70e+02  1.00e+00

# Bayesian Model Averaging
  Comparison: (fixed_H1 & random_H1) vs. (fixed_H0 & random_H0)
  Inclusion Bayes factor: 218.526 
  Inclusion posterior probability: 0.995 

# Model posterior probabilities:
          prior posterior  logml
fixed_H0   0.25  1.95e-20 -51.22
fixed_H1   0.25  2.21e-01  -7.34
random_H0  0.25  4.56e-03 -11.23
random_H1  0.25  7.74e-01  -6.09

# Posterior summary statistics of average effect size:
          mean    sd  2.5%   50% 97.5% hpd95_lower hpd95_upper  n_eff  Rhat
averaged 1.085 0.183 0.686 1.091 1.432       0.705       1.446     NA    NA
fixed    1.092 0.118 0.860 1.093 1.325       0.853       1.317 3645.6 1.001
random   1.083 0.200 0.649 1.090 1.451       0.673       1.466 3863.5 1.000
\end{verbatim}

\noindent The inclusion Bayes factor quantifying the evidence in favor of heterogeneity can be obtained using Equation~\ref{eq:BF_heterogeneity} and output from the ``Model posterior probabilities'' table.

\end{document}